

\input phyzzx.tex
\input epsf.tex

\def\ex{{\hbox{\rm e}}}

\def\tr{{\hbox{\rm Tr}}}
\def\ch{{\hbox{\rm ch}}}

\def\rl{\Lambda_{\hbox{\sevenrm R}}}
\def\wl{\Lambda_{\hbox{\sevenrm W}}}

\tolerance=500000
\overfullrule=0pt
\def\np{Nucl. Phys.}
\def\pl{Phys. Lett.}

\def\cmp{Comm. Math. Phys.}

\def\lmp{Lett. Math. Phys.}
\def\bams{Bull. AMS}
\def\am{Ann. of Math.}
\def\jpsc{J. Phys. Soc. Jap.}

\def\knot{Journal of Knot Theory and Its Ramifications}

\def\ex{{\hbox{\rm e}}}

\def\tr{{\hbox{\rm Tr}}}

\tolerance=500000
\overfullrule=0pt

\tolerance=500000
\overfullrule=0pt
 
\pubnum={US-FT/9-93\cr hep-th/9402093}
\date={December, 1993}
\pubtype={}
\titlepage

\title{THE HOMFLY POLYNOMIAL FOR TORUS LINKS FROM CHERN-SIMONS GAUGE THEORY}
\author{J.M.F. Labastida\foot{E-mail: LABASTIDA@GAES.USC.ES} and  M. Mari\~no}
\address{Departamento de F\'\i sica de Part\'\i culas\break Universidade de
Santiago\break E-15706 Santiago de Compostela, Spain}

\abstract{Polynomial invariants corresponding to the fundamental 
representation of the
gauge group $SU(N)$ are computed for arbitrary torus knots and links
in the framework of Chern-Simons gauge theory making use of knot operators. As a
result, a formula for the HOMFLY polynomial for arbitrary torus links is
presented.}

\endpage 
\pagenumber=1

\chapter{Introduction}

Chern-Simons gauge theory has shown to be a powerful tool in knot theory.
It has provided a framework in which topological invariants associated to knots,
links and graphs on compact oriented three-manifolds 
\REF\witCS{E. Witten \journal\cmp&121(89)351.}
\REF\witGR{E. Witten \journal\np&B322(89)629.}
[\witCS,\witGR] can be defined. Though a
systematic procedure to compute these topological invariants in $S^3$ exists
\REF\martin{S.P. Martin \journal\np&B338(90)244.}
[\martin],
explicit computations have been carried out only in a few cases
\REF\nos{J.M.F. Labastida and A.V. Ramallo \journal\pl&B227(89)92
\journal\pl&228(89)214, 
{\sl Nucl. Phys.} {\bf B} (Proc. Suppl.) {\bf 16B} (1990) 594.}
\REF\llr{J.M.F. Labastida, P.M. Llatas and A.V. Ramallo
\journal\np&B348(91)651.}
\REF\ygw{K. Yamagishi, M.-L. Ge and Y.-S. Wu \journal\lmp&19(90)15.}
\REF\kg{R.K. Kaul and T.R. 
Govindarajan\journal\np&B380(92)293\journal\np&B393(93)392
\journal\np&B402(93)548}
\REF\poli{J.M. Isidro, J.M.F. Labastida and A.V. Ramallo
\journal\np&B398(93)187.}
[\witCS,\witGR,\martin,\nos,\llr,\ygw,\kg,\poli]. The method proposed in
[\martin] as well as other methods based on skein rules
\REF\muk{S. Mukhi, ``Skein Relations and Braiding in Topological Gauge
Theory", Tata preprint, TIFR/TH/98-39, June 1989.}
\REF\horne{J.H. Horne \journal\np&B334(90)669.}
[\muk,\horne,\ygw,\kg] are not very useful to obtain general expressions for
sets of knots as, for example, torus knots. In this respect, the approach
proposed in [\nos,\llr] in which a knot is represented by an operator on the
Hilbert space of the corresponding Chern-Simons gauge theory seems more
promising. This method has been used in [\poli] and  a formula for the
invariants of torus knots and links carrying  arbitrary representations of the
gauge group $SU(2)$ has been presented. For the fundamental representation it
covers the case of the Jones polynomial \REF\jones{V.F.R. Jones
\journal\bams&12(85)103.} \REF\jonesAM{V.F.R. Jones\journal\am &  126  (87)
335.} [\jones,\jonesAM], while for higher dimensional representations it covers
the case of  the Akutsu-Wadati polynomials
\REF\aw{Y. Akutsu and M. Wadati
\journal\jpsc&56(87)839;\journal\jpsc&56(87)3039 
Y. Akutsu, T. Deguchi and M. Wadati
\journal\jpsc&56(87)3464;\journal\jpsc&57(88)757; for a review see
M. Wadati, T. Deguchi and Y. Akutsu, Phys. Rep. {\bf 180} (1989)247. }
[\aw].

The aim of this paper is to obtain a formula for the HOMFLY polynomial
\REF\homfly{P. Freyd, D. Yetter, J. Hoste, W.B.R. Lickorish, K. Millet and
A. Ocneanu \journal\bams&12(85)239.}
[\homfly,\jonesAM] for
arbitrary torus links. For the case of torus knots a formula 
for the HOMFLY polynomial was first presented by Jones in [\jonesAM], and
reobtained using quantum groups by  Rosso and Jones in 
\REF\rosso{M. Rosso and V. Jones\journal\knot&2(93)97} 
[\rosso]. In this paper we will present a derivation of that formula from
Chern-Simons gauge theory. This will be done by studying the vacuum expectation
value of a Wilson line with the geometry of a torus knot carrying the
fundamental representation of $SU(N)$. Our approach is based on the use of the
knot operators presented in [\nos,\llr]. The difficulties of the calculation
reside in the fact that one must work with a generic $N$. Fortunately, the
fusion rules inherent to Chern-Simons gauge theory are powerful enough to
transform sums of order $N$ into polynomials where $N$ is hidden after a
suitable choice of variables. Once the formula for torus knots is obtained we
go on and analyze the case of torus links in this
framework. As a result  we  present the corresponding formula for the
HOMFLY polynomial. This is a new result in knot theory.

The paper is organized as follows. In sect. 2 we summarize the results on
Chern-Simons gauge theory that will be needed in our computations.
In sect. 3 we calculate the HOMFLY polynomial for torus knots obtaining a
result in full agreement with previous ones. The analysis leading to the
formula for torus links is contained in sect. 4. In sect. 5 this formula is
studied in the limits $N\rightarrow 2$ and $N\rightarrow 0$ to compare its form
with known ones for the Jones polynomial and for the Alexander
polynomial, respectively. 
In sect. 6 we add final comments and remarks on our results.
The paper contains four appendices.
The first one   summarizes our group theoretical conventions. 
The second one deals with the proof of a relation used in the sect. 5.
The third one contains a computer routine written in 
{\sl Mathematica}$^{\hbox{\sevenrm TM}}$ which implements the formula obtained
in sect. 4 for the HOMFLY polynomial. The fourth and last Appendix contains a
table of polynomial invariants for some torus knots and links.

\chapter{Chern-Simons Gauge Theory}

In this section we will make a short review of previous works
[\witCS,\nos,\llr] which will lead us to state propositons 2.4 and 2.5  below
on knot operators. Let  $M$ be a boundaryless three-dimensional manifold and
let $A$ be a connection associated to a principal $G$-bundle for some Lie group
$G$. Chern-Simons gauge theory is defined by considering the action,
$$
S(A) = {1\over 4\pi} \int_M \tr\big( 
A \wedge dA + {2 \over 3} A \wedge A \wedge A \big),
\eqn\chern
$$
where $\tr$ is a bilinear non-degenerate invariant form on the Lie algebra
${\cal G}$ of $G$. Integrating this action over gauge non-equivalent
connections one obtains the partition function of the theory,
$$
Z(M) = \int [{\cal D} A]_M \exp{(ikS(A))},
\eqn\parfun
$$
where $k$ is a positive integer. Two connections are gauge-equivalent if they
are related by a gauge transformation $A \rightarrow h^{-1}A h + h^{-1}dh$
where $h$ is a mapping from $M$ to $G$. Because of the integer character of
$k$, the mappings $h$ which keep $\exp{(ikS(A))}$ invariant under gauge
transformations can be connected as well as non-connected to the identity.
The action \parfun, being the integral of a three-form, does not depend on the
metric on $M$. Therefore, the partition function \parfun\ is a function of $k$
which is a topological invariant. 

Other topological invariant quantities can be
constructed by introducing operators in the integrand of the functional
integral present in \parfun. In order to have topological invariant quantities,
these operators must be gauge invariant and metric independent. An important
class of these operators are  the Wilson line operators.  Let $\gamma$ be a
closed curve in $M$ and let $R$ be an irreducible representation of the gauge
group. The Wilson line operator  associated to $\gamma$ and $R$ is,
$$
W^{\gamma}_R(A)= \tr_R ( {\hbox{\rm P}} \exp\int_{\gamma} A),
\eqn\wline
$$
where P denotes a path-ordered product along $\gamma$. The topological
invariants correspond to the vacuum expectation value of products of Wilson
line operators: $$
{\langle W^{\gamma_1}_{R_1} \cdots W^{\gamma_n}_{R_n} \rangle }_M =
(Z(M))^{-1} \int [{\cal D}A]_M \big(\prod _{i=1}^n W^{\gamma_i}_{R_i}\big)
\,\,\ex^{iS(A)}.
\eqn\veva
$$
The functional integral in \veva\ leads to quantities which depend on the
topological properties of the embedding in the three-manifold $M$ of the link
defined by the set of curves $\gamma_i$. As first shown by Witten in [\witCS]
the quantity $\langle  \prod _{i=1}^n W^{\gamma_i}_{R_i} \rangle $, which
is a function of $k$, is related to invariant polynomials associated to links.

The expectation values \veva\ can be computed using quantum field-theoretical
techniques. Non-perturbative [\witCS,\martin,\nos,\llr,\ygw,\kg,\poli] as well
as perturbative  \REF\gmm{E. Guadagnini, M. Martellini
and M. Mintchev\journal\pl&B227(89)111 \journal\pl&B228(89)489
\journal\np&B330(90)575} \REF\alr{L. Alvarez-Gaum\'e, J.M.F. Labastida, 
and A.V. Ramallo \journal\np&B334(90)103, {\sl \np} {B} (Proc. Suppl.)
{\bf 18B} (1990) 1} 
\REF\bnw{D. Bar-Natan and E. Witten\journal\cmp&141(91)423}
\REF\unknot{M. Alvarez and J.M.F. Labastida\journal\np&B395(93)198}
[\gmm,\alr,\bnw,\unknot] methods have been applied to obtain
information about these quantities. In this paper we will carry out a
non-perturbative calculation of these observables using the operator formalism
proposed in [\nos] and further developed in [\llr,\poli].

In order to compute the vacuum expectation values \veva\ we first decompose $M$
as the joint of two manifolds $M_1$ and $M_2$ sharing a common boundary.
The joint of $M_1$ and $M_2$ is performed by identifying their boundaries via an
homeomorphism. The functional integral \veva\ is decomposed into two functional
integrals over $M_1$ and $M_2$, each of which defines a wave functional
depending on the gauge field configuration at their boundaries. Each wave
functional corresponds to a state on the Hilbert space associated to the
manifolds $M_1$ and $M_2$. If $G$ is a compact group  this Hilbert space is a
finite dimensional vector space. If $f$ is the operator representation in the
Hilbert space associated to $M_1$ of the homeomorphism corresponding to the
joint of the two manifolds $M_1$ and $M_2$, and $|\Psi_1\rangle$ and
$|\Psi_2\rangle$ represent the states associated to each of the wave
functionals, the vacuum expectation value \veva\ takes the form,
$$
{\langle \Psi _2 \vert f \vert \Psi_1 \rangle\over 
\langle \vert f \vert \rangle}.
\eqn\producto
$$
where $\vert \rangle$ represent the state corresponding to $M_1$ with no Wilson
lines inserted.

 The success of this approach resides in the ability to construct
the Hilbert spaces associated to three-dimensional manifolds with boundary.
Every boundaryless three-manifold can be represented by one of its Hegaard
splittings, \ie, every boundaryless three-manifold can be obtained as the joint
of two solid $g$-handle bodies. These are three-manifolds whose boundary is a
Riemann surface of genus $g$. Therefore one must build the
Hilbert space associated to $g$-handle bodies and the representations on it of
the homeomorphisms on their surfaces. If some Wilson lines are cut in the
splitting of the manifold $M$ the boundaries possess marked points and this
Hilbert space corresponds [\witCS] to the space of conformal blocks of an
appropriate two-dimensional conformal field theory defined 
on a Riemann surface of
genus $g$. However, if no Wilson lines are cut, there are not marked points,
and the resulting Hilbert space is related [\witCS] to the characters of the
corresponding conformal field theory. Unfortunately, not much is known in
general about these spaces. Only their dimensions are known 
\REF\verlinde{E. Verlinde\journal\np&B300(88)360}
\REF\witgt{E. Witten\journal\cmp&141(91)153\journal{J.
Geom. Phys.}&9(92)303}
[\verlinde,\witgt] for arbitrary $g$. 
However, for $g=0$ and $g=1$ with no
marked points the structure of these spaces is well known 
\REF\kz{V.I. Knizhnik and A. Zamolodchikov\journal\np&B247(84)83}
\REF\gepwit{D. Gepner and E. Witten\journal\np&278(86)493}
[\kz,\gepwit].

The connection between Chern-Simons gauge theory and conformal field theory
recasts many problems in knot theory in terms of problems in conformal field
theory. However, this connection by itself is not enough to build a framework in
which knot invariants can be computed. To be able to compute the vacuum
expectation values \veva\ one needs to have a representation of the Wilson
lines as operators acting on the corresponding Hilbert space, \ie, on the space
of conformal blocks and on the space of characters of the associated conformal
field theory. In [\nos], a program was initiated to make explicit the
connection between Chern-Simons gauge theory and conformal field theory. This
creates a framework in which the operators associated to Wilson lines can be
constructed. To compute the functional integrals leading to the wave
functionals it is necessary to know the structure of the space of
gauge-inequivalent flat connections [\nos,\llr] on Riemann surfaces of genus
$g$. Unfortunately, not much is known on these spaces (this is the Chern-Simons
counterpart of our lack of knowledge on the space of conformal blocks for
arbitrary $g$). Only their dimensions are known and in fact they have been
obtained from the associated knowledge in conformal field theory [\witgt]. For
$g=0$ with marked points and for $g=1$ with no marked points, however, these
spaces are well known and the analysis of Chern-Simons theory can be carried
out as shown in  \REF\conblock{J.M.F. Labastida and A.V.
Ramallo\journal\pl&B227(89)92}  [\conblock] and [\nos,\llr], respectively. In
the process, the operators for Wilson lines which may lie on a Riemann surface
of genus 1 without self-intersection have been constructed. They correspond
therefore to torus knots. We will refer to these operators as knot operators. 
As will be clear from the next two sections, these operators allow to compute
the vacuum expectation value \veva\ for arbitray torus links carrying arbitrary
representations lying on any lens space.

Knot operators will be the main tool of our work and we summarize their
definition and properties in the following propositions. The reader is referred
to [\llr]  for a proof of these propositions. Let us consider the gauge group
$SU(N)$, a three-manifold $M_1=T^2$ which is  a solid torus with modular
parameter $\tau$, and let us define $l=k+N$. The structure of the corresponding
Hilbert space is described as follows.

{\bf Proposition 2.1.} An orthonormal basis of the Hilbert space associated to
$T^2$ with no marked points on its surface is made up  by the states
$|p\rangle$ where $p\in {\cal F}_l$, the fundamental chamber of the weight
lattice of $SU(N)$.

The fundamental chamber ${\cal F}_l$ is made out of all linearly independent
Weyl-antisymmetric combinations of vectors on the lattice
$\wl/l\rl$, where $\wl$ and $\rl$ are the weight and root lattices of $SU(N)$
respectively. A summary of our group-theoretical conventions as well as a
description of the fundamental chamber ${\cal F}_l$ is contained in Appendix A.
We will take as representatives of the weights in ${\cal F}_l$ the ones of the
form $p=p_i\lambda_i$ with $p_i>0$ and $\sum_{i=1}^r p_i <l$, 
$\lambda_i$, $i=1,...,r$ being the fundamental weights, and $r=N-1$ the
rank of $SU(N)$. 
The form of the knot operators in this basis is stated in the
following proposition. 

{\bf Proposition 2.2.} Let the pair $(n,m)$ denote a torus knot ($n$ and $m$
are coprime integers) on the surface of $T^2$ carrying an
irreducible representation of $SU(N)$ with highest weight $\Lambda$. If $m$
denotes the number of times that the torus knot winds around the axis of the
torus, the Wilson line corresponding to that knot is represented in the basis
chosen in proposition 2.1 by the following operator:
$$
W^{(n,m)}_{\Lambda}|p\rangle=\sum_{\mu \in M_{\Lambda}}
\exp \bigl[ i\pi \mu^2 {nm\over k+N}+2\pi i {m\over k+N} 
p\cdot\mu \bigr] |p+n\mu \rangle.
\eqn\venator 
$$
In this equation $M_{\Lambda}$ is the set of weights corresponding to the
$SU(N)$ irreducible  representation of highest weight $\Lambda$. 

The operators \venator\ are called {\it knot operators}.
The inner product appearing in $\mu^2$ and $p\cdot\mu$ in these operators is the
natural one in weight space, \ie, the one corresponding to the inverse of the
Cartan matrix of $SU(N)$. Of all the states $|p\rangle$ with $p\in {\cal F}_l$,
the one corresponding  to $\rho=\sum_{i=1}^r \lambda_i$ has a
special significance since it can be regarded as the vacuum. Indeed, as proven
in [\llr], we have the following proposition.

{\bf Proposition 2.3.} Let $\Lambda$ be the highest weight of an irreducible
representation of $SU(N)$ and let $W^{(n,m)}_{\Lambda}$ be the corresponding
knot operator. Then:
$$
W^{(1,0)}_{\Lambda}|\rho\rangle = |\rho + \Lambda\rangle.
\eqn\vacio
$$

Acting on $|\rho\rangle$, the operator $W^{(1,0)}_{\Lambda}$ builds a single
vector of ${\cal F}_l$. It can therefore be regarded as a `creation operator'
which creates the state corresponding to the representation it carries when
acting on the vacuum $|\rho\rangle$. We will identify $|\rho\rangle$ with the
state which corresponds to a solid torus with no Wilson lines on it.

Once the knot operators are built, in
order to compute \producto\ one needs also the representation of the
homeomorphisms on the surface of $T^2$ in the basis of proposition 2.1. These
homeomorphisms are generated by the modular transformations $S$ and $T$ on that
surface which possess the following representations (see, for example, the
Appendix of [\llr]): $$
\eqalign{
T_{p,p'}=&\delta_{p,p'}\ex^{2\pi i(h_p-{c\over 24})},\cr
S_{p,p'}=&\Big({N\over
k+N}\Big)^{{N-1\over2}} {1\over N^{N\over2}}
(i)^{{N(N-1)\over2}}\sum_{w\in W} \epsilon(w) \exp \Bigl[{-2\pi ip\cdot
w(p')\over k+N} \Bigr],\cr}
\eqn\modular
$$
where,
$$
h_p={p^2-\rho^2 \over 2(k+N)},\,\ c={k(N^2-1)\over k+N}.
\eqn\choro
$$
In \modular\ $W$ is the Weyl group of $SU(N)$ and $\epsilon(w)$ the signature
of the Weyl transformation $w$. 

Lens spaces are boundaryless three-dimensional manifolds which can be built by
the joint of two tori. The gluing is carried out by an homeomorphism whose
representation in the Hilbert space of proposition 2.1 is written in terms of
the generators \modular\ 
\REF\dani{U. Danielsson\journal\pl&B220(89)137}
\REF\lisa{L. Jeffry\journal\cmp&147(92)563}
[\dani,\lisa]. Let us call this representation $F$. Using
\producto, the vacuum expectation value for a Wilson line corresponding to a
torus knot carrying an $SU(N)$ irreducible representation  of highest weight
$\Lambda$ lying on a lens space is,
$$
V^{(n,m)}_{\Lambda}\big\vert_F={\langle\rho\vert F
W^{(n,m)}_{\Lambda}|\rho\rangle\over
                            \langle\rho\vert F\vert\rho\rangle}.
\eqn\vevdos
$$

The quantity \vevdos\ fails to be one of the standard polynomial invariants for
three reasons. First of all, Chern-Simons gauge theory provides topological
invariants for framed knots on framed manifolds. In expression \vevdos\ a
choice of frame for the knot and  for manifold has been tacitly made.
Invariants are typically given in standard frames and we have to be sure that
such a choice is made in \vevdos, or that we correct for it.
The choice of
frame in the lens space is hidden in the choice made for $F$ in \vevdos. 
For the three-sphere, which will be the case of interest  in this paper, the
standard frame is just the one corresponding to $F=S$. The choice of frame for a
 torus knot made in \vevdos\ when $F=S$, as shown in [\llr,\poli], corresponds
to $nm$ windings. This can be easily seen thinking of the knot on the torus as a
band and analyzing the shape of this band after gluing via an $S$ modular
transformation with another  torus. The effect due to framing on knot invariants
in Chern-Simons gauge theory is well known [\witCS]. In this particular case it
accounts for a factor 
$$
\ex^{2\pi i nm h_{\rho+\Lambda}},
\eqn\framing
$$
where $h_\Lambda$ is given in \choro.
Knot invariants are usually given in the standard frame or frame in which there
are no windings. Thus we must correct \vevdos\ multiplying it by the inverse of
the framing factor \framing.

The second reason why \vevdos\ does not correspond to the standard polynomial
invariant is that the relative orientation chosen for the homology cycles on the
torus is the opposite to the standard one [\llr,\poli]. This is easily
corrected in \vevdos\ substituting $m\rightarrow -m$. The third reason is just
that the standard normalization for knot invariants is such that for the unknot
their value is 1. We must therefore normalize \vevdos\ by its value for the
unknot. Taking  all these facts into account we state the following proposition.

{\bf Proposition 2.4.} The normalized knot invariant for a torus knot $(n,m)$
in its standard framing, carrying an $SU(N)$ irreducible representation of
highest weight $\Lambda$, on $S^3$ in the standard framing, is, 
$$
X^{(n,m)}_\Lambda =  \ex^{2\pi i nm h_{\rho+\Lambda}}
{V^{(n,-m)}_{\Lambda}\big\vert_S \over V^{(1,0)}_{\Lambda}\big\vert_S} 
=\ex^{2\pi i nm h_{\rho+\Lambda}}{\langle\rho\vert S
W^{(n,-m)}_{\Lambda}|\rho\rangle \over \langle\rho\vert S
W^{(1,0)}_{\Lambda} \vert\rho\rangle }.
\eqn\equis
$$

The purpose of this paper is first to compute \equis\ for the case of the
fundamental representation of $SU(N)$. For a given torus knot $(n,m)$, the
invariant given in \equis\ is a function of $k$ and $N$. From \venator,
\modular\ and \choro\ it is clear that the dependence on $k$ always enters as
$\exp(2\pi i/(k+N))$ and therefore we can regard  $X^{(n,m)}_\Lambda $ as a
function of $t$ and $N$ where, $$
t=\ex^{2\pi i \over k+N}.
\eqn\late
$$
When $\Lambda = \lambda_1$ it turns out that the rest of the dependence on $N$
in \equis\ can be written in terms of the variable
$$
\lambda = t^{N-1}.
\eqn\lalanda
$$
This is not obvious at all from \equis\ since, as can be seen from \venator\
and \modular, sums whose ranks depend on $N$ enter in the computation of \equis.
As shown in the next section this occurs independently of the value of $N$ as
long as $N>n$. The result is a rational function of $t$ and $\lambda$  which
agrees with the expression for the HOMFLY polynomial for torus knots given in
[\jonesAM].

In sect. 4 we will use \equis\ to compute invariants for torus links. A
torus link $(n,m)$, where $n$ and $m$ are two arbitrary integers, possesses $s$
components, where $s$ is the greatest common divisor of $n$ and $m$. It can be
built by a successive insertion of its components, which are $(n/s,m/s)$ torus
knots, into the interior of a torus $T^2$. This implies that in the
corresponding invariant 
 a product of knot operators will enter. As discussed in  [\poli], the adequate
deframing factor to contact with results in the standard framing is the same as
in \equis. The following proposition was proved in [\poli]:

{\bf Proposition 2.5.} The normalized knot invariant for a torus link $(n,m)$
in its standard framing, whose components carry the same  $SU(N)$ irreducible
representation of highest weight $\Lambda$, on $S^3$ in the standard framing,
is,  $$
X^{(n,m)}_\Lambda 
=\ex^{2\pi i nm h_{\rho+\Lambda}}{\langle\rho\vert S
\Big( W^{(n/s,-m/s)}_{\Lambda}\Big)^s|\rho\rangle \over \langle\rho\vert S
W^{(1,0)}_{\Lambda} \vert\rho\rangle },
\eqn\equisdos
$$
where $s$ is the greatest common divisor of $n$ and $m$.

As for the case of torus
knots, this invariant turns out to be a function of $t$ and $\lambda$ as
defined in \late\ and \lalanda\ (actually a rational
function of $\sqrt{t}$ and $\sqrt{\lambda}$). The evaluation of \equisdos\ for
the fundamental representation of $SU(N)$ will provide an expression for the
HOMFLY polynomial for torus links. Its computation, which is the main result of
this paper, will be carried out in sect. 4.

\endpage

\chapter{HOMFLY polynomial for torus knots}

In this section we will make use of proposition 2.4 to compute the HOMFLY
polynomial for torus knots. We must evaluate \equis\ for the fundamental
representation of $SU(N)$, \ie, we must take $\Lambda=\lambda_1$. The result is
stated in the following theorem:

{\bf Theorem 3.1.} The HOMFLY polynomial for a torus knot $(n,m)$ is given
by:
$$
X^{(n,m)}_{\lambda_1} = \Big ( {1-t \over 1-t^n }\Big)
{\lambda^{{1\over2}(m-1)(n-1)} \over \lambda t-1} \sum_{p+i+1=n \atop p, i
\ge 0} (-1)^i t^{mi + {1 \over 2} p(p+1)} {\prod_{j=-p}^i (\lambda t-t^j)
\over (i)! (p)!}
\eqn\nuria
$$
where $(x)=t^x-1$, $\lambda= t^{N-1}$ and $t=\exp(2\pi i/(k+N))$.

{\bf Proof.} The rest of this section deals with the proof of this theorem. 
First notice that since an $(n,m)$ torus knot is isotopically equivalent to the
$(-n,-m)$ torus knot, we can restrict ourselves to torus knots with $n>0$.
Also, as stated in sect. 2, we will consider the case in which $N>n$.
Let us begin
working out the action of the knot operator  $W^{(n,m)}_{\lambda_1}$ on the
vacuum state. Using \venator\ and taking into account the scalar products (A20)
and (A21), we have: 
$$
 W^{(n,m)}_{\lambda_1}|\rho\rangle=\sum^{N}_{i=1}\exp[i\pi(1-{1\over
N}){nm\over k+N}+2\pi i {m\over k+N}{1\over 2}(N-2i+1)]|\rho+n\mu_i\rangle,
\eqn\elizeth
$$
where $\mu_i$, $i=1,...,N$, are the weights in $M_{\lambda_1}$ whose 
explicit expression is given in (A.17).
Following proposition 2.1, we must find the canonical representatives in the
fundamental chamber ${\cal F}_l$ of the weights appearing in this sum. These
weights have the following structure: 
$$ 
\eqalign{
 &(n+1,1,\cdots,1),\,\,\,\,\ i=1\cr
 &(1-n,1+n,1,\cdots,1),\,\,\,\, i=2\cr
 &\,\,\,\,\,\,\,\,\,\,\,\,\,\,\ \vdots \cr
 & (1,\cdots,1,1-n),\,\,\,\,\,\ i=N \cr}
\eqn\apple
$$
Every weight in the weight lattice can be written as
$w(\mu)+l\alpha$, where $w$ is an element of the Weyl group, $\alpha$ a
root, and $\mu$ is a weight whose components are non-negative. In
the Hilbert space described in proposition 2.1 the weights
which possess one or more vanishing components are represented by
null vectors. Since $N>n$ there is no need to add terms of the form $l\alpha$
to the weights in 
\apple\ to bring them to a form in which their components are non-negative. 
A series of Weyl reflections will be sufficient.
If $n=1$ all the weights in \apple\ except the first one have one vanishing
component and therefore there is only one contribution in the sum present in
\elizeth. If $n>1$, notice first that the first weight
in \apple, $\rho+n\mu_1$, is already in ${\cal F}_l$. For the rest we have the
following cases:

1) $2\le i\le n$. We perform the chain of Weyl reflections: $$
\eqalign {
 &(1,\cdots,1,1-n,{\buildrel i \over {1+n}},1,\cdots,1)\cr
&\,\,\,\,\,\,\,\,\,\,\,\,\,\,\,\,\,\,\,\,\,\,\,\,\,\,\ \downarrow
\sigma_{i-1}\cr    
 &(1,\cdots,1,2-n,n-1,{ \buildrel i \over 2},1,\cdots,1)\cr
&\,\,\,\,\,\,\,\,\,\,\,\,\,\,\,\,\,\,\,\,\,\,\,\,\,\,\ \downarrow
\sigma_{i-2}\cr 
&\,\,\,\,\,\,\,\,\,\,\,\,\,\,\,\,\,\,\,\,\,\,\,\,\,\,\,\ \vdots \cr
&\,\,\,\,\,\,\,\,\,\,\,\,\,\,\,\,\,\,\,\,\,\,\,\,\,\,\,\ \downarrow
\sigma_1\cr
 & (n+1-i,1,\cdots,1,{ \buildrel i \over 2},1,\cdots,1)\cr}
\eqn\casandra
$$
After $i-1$ Weyl reflections, we obtain the following  weight in ${\cal F}_l$:
$$
|\rho+n\mu_i\rangle=(-1)^{i-1}|\rho+(n-i)\lambda_1 +
\lambda_i\rangle,\,\,\,\,\,\ 2\le i \le n,\,\,\,\,\,\,\,\,\, n>1.
\eqn\cesaria
$$

2) $i>n$. The chain of Weyl reflections is like the one in \casandra:
 $$
  \eqalign{
  &(1,\cdots,1,1-n,{ \buildrel i \over {1+n}},1,\cdots,1)\cr
&\,\,\,\,\,\,\,\,\,\,\,\,\,\,\,\,\,\,\,\,\,\,\ \downarrow
\sigma_{i-1}\cr 
  &(1,\cdots,1,2-n,n-1,{ \buildrel i \over 2},1,\cdots,1)\cr
&\,\,\,\,\,\,\,\,\,\,\,\,\,\,\,\,\,\,\,\,\,\,\,\ \downarrow
\sigma_{i-2}\cr 
&\,\,\,\,\,\,\,\,\,\,\,\,\,\,\,\,\,\,\,\,\,\,\,\,\,\,\,\ \vdots \cr
&\,\,\,\,\,\,\,\,\,\,\,\,\,\,\,\,\,\,\,\,\,\,\,\,\ \downarrow
\sigma_{i-n-1}\cr
  &(1,\cdots,1,{\buildrel i-n \over 0},1,\cdots,1,{ \buildrel i
\over 2},1,\cdots,1)\cr} 
\eqn\eliane
$$
After $n-1$ reflections the weights get a vanishing component and therefore all
these weights correspond to null vectors and do not contribute to the sum in
\elizeth. This fact is very important in this calculation because it implies
that the sum \elizeth\ is truncated. Its upper limit turns out to be $n$
instead of $N$. A consequence of this (and other similar phenomena which will
appear below) is that, as emphasized at the end of the last section, the knot
invariant will hide all its dependence on $N$ through the variables $t$ and
$\lambda$ defined in \late\ and
\lalanda. Using these variables, the sum in \elizeth\ becomes,
$$
W^{(n,m)}_{\lambda_1}|\rho\rangle=t^{mn{N-1\over2N}+m{N+1\over2}}
\sum_{i=1}^n (-1)^{i-1} t^{-mi} |\rho+ (n-i)\lambda_1+\lambda_i\rangle,
\eqn\chloe
$$
where the factor $(-1)^{i-1}$ is due to the Weyl reflections carried out. This
equation is valid for any $n\geq 1$ as long as $N>n$.
The vacuum expectation value \vevdos\ which enters \equis\ takes the form:
$$
V^{(n,m)}_{\lambda_1} = {\langle\rho|S W^{(n,m)}_{\lambda
_1}|\rho\rangle \over \langle\rho |S|\rho\rangle }  
 =t^{nm{N-1\over2N}+m{N+1\over 2} }  \sum_{i=1}^n
(-1)^{i-1} t^{-mi}  { S_{\rho,\rho+(n-i)\lambda_1+\lambda_i}
\over S_{\rho,\rho}}.
 \eqn\isabelle
$$
If $\Lambda$ is a highest weight, the ratio 
$S_{\rho,\rho+\Lambda}/S_{\rho,\rho}$ can be written in terms of the character
associated to $\Lambda$ with the help of the Weyl formula:
$$
{S_{\rho,\rho + \Lambda}\over S_{\rho,\rho}}={ \sum_{w \in
W} \epsilon(w) \exp[-{2\pi i\over k+N} \rho \cdot w(\rho+\Lambda)] \over 
\sum_{w \in W} \epsilon (w) \exp[-{2\pi i\over k+N} \rho \cdot w(\rho)] }={\rm
ch}_{\Lambda} [-{2\pi i \over k+N} \rho].
\eqn\cecilia
$$
All the weights entering \isabelle\ can be thought of as highest weights and
therefore we can express $V^{(n,m)}_{\lambda_1}$ in terms of characters:
$$ 
V^{(n,m)}_{\lambda_1}=t^{nm{N-1\over 2N}+m{N+1 \over 2}}  \sum_{i=1}^n
(-1)^{i-1} t^{-mi} {\rm ch}_{(n-i)\lambda_1 +\lambda_i} [-{2\pi i\over
k+N}\rho].
\eqn\eloisa
$$

Let us compute first $V^{(1,0)}_{\lambda_1}$, which is the quantity entering the
denominator in \equis. From \isabelle\ and \cecilia\ follows that one needs
to compute the character for the fundamental representation. This calculation
is done very simply just summing over the weights of the representation:
$$
{\rm ch}_{\lambda_1}
[-{2\pi i \over k+N}] = \sum_{\mu \in M_{\lambda_1}} t^{-\mu \cdot
\rho}=\sum_{j=1}^N t^{-{1\over2}(N-2j+1)}=t^{-{1\over2}(N-1) }{t^N-1 \over t-1}.
\eqn\manzana
$$
Using this result, it turns out that
$$
V^{(1,0)}_{\lambda_1}= \lambda^{-{1\over 2}}{\lambda t-1\over t-1},
\eqn\judith
$$
which has been written entirely in terms of the variables $\lambda$ and $t$ in
\late\ and \lalanda.

For representations different from the fundamental one, however, it is more
useful to compute the character using its expression in terms of a product over
positive roots:
$$
{\rm ch}_{\Lambda}
[-{2\pi i \over k+N}] =
\sum_{\mu \in M_\Lambda} t^{-\mu \cdot \rho} = \prod_{\alpha>0}{t^{{1\over
2}\alpha \cdot (\rho +  \Lambda)}-t^{-{1\over 2}\alpha \cdot
(\rho+\Lambda)}\over t^{{1\over 2} \alpha \cdot \rho}-t^{-{1\over 2}\alpha
\cdot \rho}}. \eqn\luiza
$$
In this equation, the symbol $\alpha > 0$ indicates that the product has to be
performed over all the positive roots. For $SU(N)$ these are given in Appendix
A. Our next task is to compute the characters appearing in \eloisa\ with the
help of this formula.

In order to simplify our notation, from now on we will denote
${\rm ch}_{\Lambda} [-2\pi i / (k+N)$ simply by ${\rm ch}_{\Lambda}$. Also, we
introduce  the following notation regarding   $q$-numbers,  $q$-factorials
and
$q$-combinatorial numbers:
$$
\eqalign{               
[x] =& t^{x \over 2} - t^{-{x\over 2}},\cr
[x]!=&[x] [x-1]\cdots [1],\cr
\Bigg [ {x \atop y} \Bigg]=&{[x]!\over [x-y]! [y]!}.\cr}
$$
This allows us to write the character formula in the form:
$$
{\rm ch}_\Lambda= \prod_{\alpha>0}{[\alpha \cdot (\rho +\Lambda)]\over
[\alpha \cdot  \rho]} 
\eqn\amelia
$$

In order to compute \eloisa\ we must perform the products in \amelia\ for
weights of the form $p\lambda_1 +\lambda_i$, where $p=n-i$. Taking into account
the form of the positive roots  listed in Appendix A, this suggests to organize
the product in \amelia\ splitting the set of positive roots into three groups.
Let us carry out the computation of the character for each group:

1) Positive roots beginning with $\alpha_1$ and ending before $\alpha_{i}$. We
label them by the index $k$ as follows:
$$
\eqalign{
&\alpha_1,\,\,\,\,\,\,\,\,\,\,\,\,\,\,\,\,\,\ k=1,\cr
&\alpha_1 + \alpha_2,\,\,\,\,\,\,\,\,\ k=2, \cr
&\,\ \vdots \cr
&\alpha_1+ \cdots + \alpha_{i-1},\,\,\ k=i-1.\cr}
$$
The scalar products entering \eloisa\ are:
$$
\eqalign{
&\Lambda \cdot \alpha = p,\cr
& \rho\cdot\alpha =k,\cr}
$$
and we have the contribution:
$$
\prod_{k=1}^{i-1}{[p+k]\over [k]}={[p+i-1]!\over[p]![i-1]!}.
\eqn\pla
$$

2) Positive roots beginning with $\alpha_1$ and ending after $\alpha_i$.
Similarly, these roots are labeled by $k$ in the following way:
$$
\eqalign{
&\alpha_1+\cdots+\alpha_{i},\,\,\,\,\ k=i,\cr
&\alpha_1+\cdots+\alpha_{i+1},\,\,\,\,\ k=i+1,\cr
&\,\,\,\,\,\,\,\,\,\,\,\ \vdots\cr
& \alpha_1+\cdots+\alpha_{N-1},\,\,\,\ k=N-1,\cr}
$$
The scalar products are:
$$
\eqalign{
&\Lambda \cdot \alpha=p+1,\cr
&\rho \cdot \alpha=k,\cr}
$$
and the corresponding contribution is:
$$
\prod_{k=i}^{N-1}{[p+1+k]\over [k]}={[p+N]!\over
[p+i]!}{[i-1]!\over[N-1]!}.
\eqn\tano
$$

3) Positive roots which have the generic form: 
$$
\alpha_j+\cdots+\alpha_{i}+\cdots+\alpha_{k-1},\,\,\,\,\  j>1,\,\ k>i.
$$
The scalar products are,
$$
\eqalign{
&\Lambda\cdot\alpha=1,\cr
&\rho\cdot\alpha=k-j,\cr}
$$
and the corresponding contribution becomes:
$$
\eqalign{
\prod_{j=2}^{i}\prod_{k= i+1}^N {[k-j+1] \over
[k-j]}=&\prod_{j=2}^{i}  {[N+1-j]\over [N-j]}{[N-j]\over [N-j-1]}\cdots
{[i+2 -j]\over [i+1-j]}  \cr
=&\prod_{j=2}^{i}{[N+1-j] \over
[i+1-j]}={[N-1]!\over[N-i]![i-1]!}.\cr}
\eqn\banana
$$
Notice that the product whose upper limit is $N$ becomes truncated after the
cancellation of contributions from the numerator and from the denominator. 
This fact
is important because it will allow to hide all the dependence on $N$ into the
variables $\lambda$ and $t$ defined in \lalanda\ and \late.

Taking into account \pla, \tano\ and \banana, we finally
obtain a formula for the character in terms of 
$q$-numbers:
$$
{\rm ch}_{p\lambda_1+\lambda _{i}}={[i] \over [p+i]} \Bigg [
{N+p \atop  p} \Bigg] \Bigg[{N \atop i} \Bigg ].
\eqn \donimo
$$          
From this it is straightforward to write an expression for \equis\ involving
only the variables $t$ and $\lambda$. To compare with Jones' result
[\jonesAM,\rosso], it is convenient first to realize that
$$
(x) = t^{x}-1 = t^{-{x\over 2}} [x],
\eqn\penelope
$$
and therefore,
$$
\eqalign{
[x]!=&t^{-{1\over 4}x(x+1)} (x)!,\cr
\Bigg [{N+p \atop  p} \Bigg]=&t^{-{1\over2}Np} {(N+p)! \over (p)!
(N)!}= t^{-{1\over 2}Np +{1  \over 2}p(p+1)}  {\prod_{j=-p}^{-1}
(t^N -t^j) \over (p)!},\cr
\Bigg[{N \atop i} \Bigg ]=&t^{-{1 \over2}Ni +{1 \over 2}{i}^2}{(N)!
\over (N-i-1)! (i+1)!}
=t^{-{1 \over 2}Ni + {1 \over 2}i}
{\prod_{j=0}^{i-1} (t^N-t^j) \over (i)!}.\cr}  
\eqn\karina
$$

In terms of these new quantities the character is:
$$
{\rm ch}_{p\lambda_1 +\lambda_{i}}=
{\lambda^{-{n\over2}}t^{{1\over2}p(p+1)}\over t^n-1}{\prod_{j=-p}^{i-1}
(\lambda  t-t^j) \over (i-1)! (p)!}, \,\,\,\,\  n=p+i,
\eqn\marisa
$$ 
which gives the normalized vacuum expectation value:
$$
V_{\lambda_1}^{(n,m)}={\lambda^{mn\over 2N}\lambda^{m-n
\over2}\over t^n-1}\sum_{p+i+1=n \atop  p, i \ge 0}(-1)^i t^{-mi+{1 \over
2} p(p+1) }{\prod _{j=-p}^i (\lambda t-t^j) \over (i)! (p)!},
\eqn \elena
$$

It remains only to obtain the deframing phase factor. The conformal weight for
the fundamental representation of $SU(N)$ is given by \choro:
$$
h_{\rho+\lambda_1}={(\rho+\lambda_1)^2-\rho^2 \over 2(k+N)}= {N-{1 \over N}
\over 2(k+N)},
\eqn\manza
$$
which gives the deframing factor:
$$
\ex^{2\pi i nm h_{\rho+\lambda_1}} = t^{{mn \over 2}(N-{1 \over N})}.
\eqn\pera
$$
From \elena\ and \pera\ one obtains the final expression for the knot
invariant \equis, which equals the one stated in theorem 3.1.
This ends its proof.

\endpage

\chapter{HOMFLY polynomial for torus links}

In this section we will make use of proposition 2.5 to obtain a formula for
the HOMFLY polynomial for torus links. Before stating the main theorem of
this section, which is the one containing such a formula, we will analyze
equation \equisdos\ to organize its computation. The operators acting on
$|\rho\rangle$ in \equisdos\ are all the same and therefore they commute.
Using \elizeth\ it turns out that the action of this product can be written as,
$$
\eqalign{
(W_{\lambda_1}^{({n \over s},{m \over
s})})^s & |\rho \rangle =\cr 
& \hbox{\hskip-20pt}\sum_{\mu_{i_1}\in
M_{\lambda_1}\atop{\vbox{\vglue-100pt}\vdots \atop\mu_{i_s} \in M_{\lambda_1}}} 
\hbox{\hskip-10pt}\exp \biggl[ \pi i \Big (\sum_ {j=1}^s \mu_{i_j} \Big)^2 {nm
\over k+N} + 2\pi i {m \over k+N} \rho \cdot \Big (\sum_{j=1}^s \mu_{i_j} \Big
) \biggr] |\rho + {n \over s}\sum_{j=1}^s \mu_{i_j} \rangle.\cr}
\eqn\dora
$$
The weights appearing on the right hand of this expression are sums of $s$
elements from the set $M_{\lambda_1}$, so there are many configurations of 
$\mu_{i_j}$ giving the same sum $\sum_{j=1}^s \mu_{i_j}$. To take this
degeneracy into account, it is useful to introduce the set of ordered
partitions of $s$, \ie, the set of ordered sequences  $(k_{\lambda})$ verifying
$\sum_{\lambda} k_{\lambda} = s$. For each sequence, we will denote its cardinal
generally by $r$. As we prove in Appendix B, there are
$2^{s-1}$ ordered partitions of $s$. The degeneracy associated to each ordered
partition is: 
$$  
{s! \over \prod_{\lambda} k_{\lambda}!},
$$ 
and we can write the weights in the ket of \dora\ as: 
$$ 
\rho + {n \over s}(k_1 \mu_{i_1} + \cdots + k_r
\mu_{i_r}),\,\,\,\,\ i_1 <i_2 < \cdots <i_r. 
$$ 
In this way the sum over
weights decomposes into a sum over partitions and a sum over the indices 
$i_1 < \cdots <i_r$: 
$$ 
\sum_{\mu_{i_1}, \cdots , \mu_{i_s} } \cdots =
\sum_{(k_{\lambda})} {s! \over \prod_{\lambda} k_{\lambda}!} \sum_{i_1 < \cdots
<i_r} \cdots 
$$ 
Recall from the previous section that not all these weights
contribute to the sum. The general strategy is the following: apply Weyl
reflections in such a  way  that the weight becomes either one of the
weights in ${\cal F}_l$, or one which is equivalent to a null state. This last
situation occurs when a vanishing component appears in the process.
The result of applying this procedure is stated
in the following theorem.

{\bf Theorem 4.1.} Given an ordered partition of $s$, $(k_{\lambda})$, with
cardinal $r$, and a weight of the form:
$$
\rho + {n \over s}(k_1 \mu_{i_1} + \cdots + k_r \mu_{i_r}),\,\,\,\,\ i_1
<i_2 < \cdots <i_r,
\eqn\lima
$$ 
the arrangement of indices ${i_{\lambda}}$, ${\lambda=1,\cdots,r},$ 
producing a weight
in ${\cal F}_l$ is contained in the set specified by the following conditions: 
$$ 
\eqalign{ 
&({\rm I})\,\ i_{\lambda} \le {k_{\lambda} n \over s},\cr
&({\rm II})\,\ i_{\lambda}=i_{\mu}+{k_{\lambda} n \over s},\,\,\,\ 
\mu<\lambda, \cr}
$$
in such a way that, in (II), no previous index $i_{\nu}$,
$\nu<\lambda$, has the form $i_{\nu}=i_{\mu}+{k_{\nu} n \over s}$,
$\mu<\nu$.
Given an arrangement of indices like this, with $r-k$ indices verifying
condition (I) (which will be called of type I) and $k$ indices verifying
condition (II) (which will be called of type II), a weight belonging to ${\cal
F}_l$
 is obtained if and only if:  
$$
 i_{\mu}
-i_{\nu}+{k_{\nu}-k_{\mu} \over s}n \not= 0,
$$  
for every pair of indices $i_{\mu}$, $i_{\nu}$,  verifying (I).
The set of arrangements of indices selected in this way will be denoted by
${\cal I}_{(k_\lambda)}$.

To each arrangement of indices in ${\cal I}_{(k_\lambda)}$, we will
associate a  canonical representative in ${\cal F}_l$ accompanied
by a sign which corresponds to the weight in
\lima. This association is carried out by the following
procedure:
 
1) For indices of type I, which will be denoted by
$i_{\lambda_1}, \cdots,i_{\lambda_{r-k}}$, one defines a total order relation
according to: 
$$
i_{\lambda_p} \succ i_{\lambda_q}\,\ {\rm iff} \,\
i_{\lambda_p}-i_{\lambda_q}+{k_{\lambda_q}-k_{\lambda_p} \over
s}n>0.
$$
This relation defines a permutation $\tau$ of the set of indices
of type I under consideration with respect to their natural ordering: 
$$
\tau = \left(
\matrix{i_{\lambda_1}&i_{\lambda_2}&\cdots&i_{\lambda_{r-k}}\cr
i_{\tau(\lambda_1)}&i_{\tau(\lambda_2)}&\cdots&i_{\tau(\lambda_{r-k})}
\cr}\right).
$$

2) For the $k$ indices of type II, $i_{
\nu_1}, \cdots , i_{ \nu_k}, \,\ i_{ \nu_1}<
\cdots < i_{ \nu_k}$, one takes the set of indices
$i_{\hat \nu_1}, \cdots , i_{\hat \nu_k}$, verifying $i_{
\nu_p}=i_{\hat \nu_p}+{k_{\nu_p}n \over s}$, and defines on it the order
relation inherited from the natural ordering of the indices $i_{\nu_p}$:
$$
 i_{\hat \nu_p} \succ i_{\hat \nu_q}\,\ {\rm iff} \,\
i_{\nu_p}>i_{\nu_q}.
$$
This gives again a permutation $\sigma$ with respect to the natural
ordering of the set $i_{\hat \nu_p}$: 
$$
\sigma=\left( \matrix {i_{\sigma^{-1}({\hat\nu_1})}&i_{\sigma^{-1}({\hat
\nu_2})}& \cdots &i_{\sigma^{-1}({\hat\nu_k})}\cr
                         i_{\hat \nu_1}& i_{\hat \nu_2}&\cdots &
i_{\hat \nu_k}\cr}\right),
$$
with $i_{\sigma^{-1}({\hat\nu_1})}<i_{\sigma^{-1}
({\hat\nu_2})}<\cdots
<i_{\sigma^{-1}({\hat\nu_k})}$.

3) Define $r-k$ numbers $\xi(\lambda_p)$, $p=1,\cdots,r-k$,
associated to type I indices as follows:  $\xi(\lambda_p)$ is the
number of type II indices preceding the type I index $i_{\lambda_p}$
in the original arrangement of indices  
$i_1,\cdots, i_r$, in \lima.

The canonical representative in ${\cal F}_l$ of the weight we began with in
\lima\ is the weight: 
$$ 
\rho +  p_1 \lambda_1 + p_2 \lambda_2 + \cdots
+p_{r-k}\lambda_{r-k}+\lambda_{i_{\mu_1} +r-k-1}
+ \lambda_{i_{\mu_2}+r-k-2}+ \cdots +\lambda_{i_{\mu_{r-k}}},
\eqn\elpeso
$$
where $p_i$, $i=1,\cdots,r-k$, are given by:
$$
\eqalign{
p_1
=&i_{\tau(\lambda_2)}-i_{\tau(\lambda_1)}+{k_{\tau(\lambda_1)}-
k_{\tau(\lambda_2)} \over s}n - 1, \cr
p_2 =& i_{\tau(\lambda_3)} -i_{\tau(\lambda_2)} +
{k_{\tau(\lambda_2)}-k_{\tau(\lambda_3)} \over
s}n -1, \cr
&\,\,\,\,\,\,\,\,\,\,\,\,\ \vdots \cr
p_{r-k}=&{k_{\tau(\lambda_{r-k})} \over s}n  -i_{\tau(\lambda_{r-k})},
\cr}
\eqn\laspes
 $$
and the indices $i_{\mu_p}$ in \elpeso\ are the complementary ones to
the indices $\{ i_{\hat
\nu_p} \}_{p=1, \cdots, k}$, \ie, those indices $i_{\mu_p}$
such that no index $i_{\nu}>i_{\mu_p}$ has the form $i_{\nu}=i_{\mu_p}+
{k_{\nu} n
\over s}$. They are ordered according to their natural ordering: $i_{\mu_1}<
\cdots <i_{\mu_{r-k}}$. 
 
Finally, the sign associated to this weight because of the Weyl reflections
needed to obtain it is:  $$
\epsilon (\tau) \epsilon (\sigma) (-1)^{\sum_{p=1}^{r-k}
i_{\mu_p}-\mu_p +  \xi(\lambda_p)}.
\eqn\sigsig
$$

{\bf Proof.} The proof is lengthy but straightforward. 
We first analyze a few cases to obtain later the general pattern.
Let us consider a partition $(k_\lambda)$ of $s$ and 
an arrangement of indices, $i_1,\cdots,i_r$, such that the weight in
\lima\ takes the form
$$ 
(1, \cdots
,1,1-{k_1 n \over s},\buildrel i_1  \over {1+{k_1 n \over s} } ,1, \cdots
,1,1-{k_r n \over s},\buildrel i_r \over  {1+{k_r n \over s} } ,1, \cdots ),
$$
\ie, $i_1>1,i_2>i_1+1, \cdots,i_{r}>i_{r-1}+1$. The cases $i_1=1$ and
$i_{\lambda}=i_{\lambda -1}+1$ for some $\lambda$ are  analogous and give the
same pattern. This is easily seen if we write all the weights as linear
combinations of the fundamental weights $\lambda_i$. We will assume  that
$n>0$, as we did in the previous section. First notice that if ${k_1n \over s}
=1$, the corresponding state is null. Hence, let us assume that ${k_1n \over
s}>1$. Performing a chain of Weyl reflections $\sigma_{i_1-1},\cdots$, it
turns out that $i_1-2$ units are added to
$1-{k_1 n \over s}$. In carrying this out one ends in one of the following two
possibilities:

1) $1-{k_1 n \over s}+i_1-2 \ge 0$,
 
2) $1-{k_1 n \over s}+i_1-2<0$.

In the first case, \ie, $i_1>{k_1 n \over s}$, one obtains a null state. In
the second case, after $i_1-1$ Weyl reflections, the following weight is
obtained: 
$$
 ({k_1 n \over s}+1-i_1,1, \cdots,1,\buildrel i_1 \over
2,1,\cdots,1,1-{k_2 n \over s}, \cdots). 
$$
As we assumed $i_1>1$, a weight with one vanishing component is obtained if: 
$$
i_1>{k_1 n \over s}.
$$
Let us analyze what happens with the second index, $i_2$. Again, if ${k_2 n
\over s}=1$, we have a null state. If ${k_2 n \over s}>1$, we apply the chain
of Weyl reflections $\sigma_{i_2-1}, \cdots $. 
Just before applying
$\sigma_{i_1}$, 
one has added $i_2-i_1-2$ units to $1-{k_2 n \over s}$. Starting with, 
$$
(\cdots, 1,{\buildrel i_1  \over 2},1, \cdots,1,1-{k_2 n \over s},\buildrel i_2
\over {1+{k_2 n \over s}}, \cdots ),
$$ 
there are two cases:

1) $1-{k_2 n \over s}+i_2-i_1-2 \ge 0$,

2) $1-{k_2 n \over s}+i_2-i_1-2<0$.

In the first case one obtains, after $i_2-i_1-2$ Weyl reflections:
$$
( \cdots ,1, \cdots ,1,\buildrel i_1 \over 2 ,i_2-i_1-1-{k_2 n
\over s},{k_2 n \over s}-(i_2-i_1-2),1, \cdots).
$$

When one applies $\sigma_{i_1+1}$ there are three subcases:

1.1) $i_2-i_1+1-{k_2 n \over s}>0$,
 
1.2) $i_2-i_1+1-{k_2 n \over s}=0$,

1.3) $i_2-i_1+1-{k_2 n \over s}>0$.

In the first subcase, provided that $1-{k_2 n \over s}i_2-i_1-2<0$, one
obtains:
$$
i_2= {k_2 n \over s}+i_1,
$$
corresponding to the weight:
$$
( \cdots, 1,{\buildrel i_1 \over 1},1, \cdots ,1,{\buildrel i_2
\over 2}1, \cdots ).
$$
In the second subcase one arrives to a null state. In the third subcase 
the weight: 
$$
({k_1 n \over s}+1-i_1,1, \cdots,1,\buildrel i_1 \over
{i_2-i_1+1-{k_2 n \over s}},{k_2 n \over s}+1+i_1-i_2,1,
\cdots,1,{\buildrel i_2 \over 2},1, \cdots),
$$
is obtained. We can sum $i_1-2$ units to $i_2-i_1+1-{k_2 n \over s}$,
so we have two possibilities:

1.3.1) $i_2-1-{k_2 n \over s} \ge 0$,

1.3.2) $i_2-1-{k_2 n \over s}<0$.

In the first possibility,
$$ 
i_2>{k_2 n \over s},\,\ i_2 \not= {k_2 n \over s}+i_1,
$$
and we have a null state. In the second possibility, after $i_1-2$
Weyl reflections, we obtain the weight: 
$$ 
({k_1 n \over
s}+1-i_1,i_2-1-{k_2 n \over s},{ k_2 n \over s}-i_2+2,1,
\cdots,1,\buildrel i_1+1 \over 2,1,
\cdots,1,{\buildrel i_2 \over 2},1, \cdots).
$$
Now, applying $\sigma_2$, we arrive to:
$$
(i_2-i_1+{k_1-k_2 \over s}n,{k_2 n \over s}+1-i_2,1, \cdots,1,{
\buildrel i_1+1 \over 2},1, \cdots,1,{\buildrel i_2 \over 2},1,
\cdots). 
$$
Notice that in general $k_1 \not= k_2$, so the sign of the first component
depends on the partition $(k_{\lambda})$. We have three
 possible situations:

a) If $i_2-i_1+{k_1-k_2 \over s}n>0$, the first and second components
of this weight are positive. We have made $(-1)^{i_1-1+i_2-2}$ Weyl
reflections to obtain it ($i_1-1$ associated to the first component
and $i_2-i_1-2+i_1$ to the second one).

b) If $i_2-i_1+{k_1-k_2 \over s}n=0$ we have a null state.

c) If $i_2-i_1+{k_1-k_2 \over s}n<0$, we must apply again the Weyl
reflection $\sigma_1$, obtaining in this way: $$
(i_1-i_2+{k_2-k_1 \over s}n,{k_1 n \over s}+1-i_1,1, \cdots,1,{
\buildrel i_1+1 \over 2},1, \cdots,1,{\buildrel i_2 \over 2},1,
\cdots).
$$
The conclusion of this analysis is that, at this level, the {\it necessary}
conditions not to obtain a null state are the following :  $$
\eqalign{
i_2 =& i_1 +{k_2 n \over s}, \cr
i_2 &\le {k_2 n \over s}. \cr}
$$
This procedure seems rather systematic and it is easy to infer the
general pattern. However, a basic new ingredient appears
only when considering the third index $i_3$ and successive indices.
For $i_3$ we have two very different cases depending on the values of
the second index. Suppose first  $i_2 \le {k_2 n \over s}$. The
weight is:
$$
( \cdots ,1,{\buildrel i_1+1 \over 2},1, \cdots,1,{\buildrel i_2
\over 2},1,1-{k_3 n \over s},\buildrel i_3 \over {1+{k_3 n
\over s}},1, \cdots).
$$
Again, if $i_3-i_2 \ge 1+{k_3 n \over s}$ we obtain a null state.
 If not, we have the possibility: 
$$
i_3=i_2+{k_3 n \over s},
$$
corresponding to the weight:
$$
(\cdots, 1,{\buildrel i_1+1 \over 2},1,
\cdots,1,{\buildrel i_2 \over 1},1,\cdots ,1,{\buildrel i_3
\over 2},1, \cdots ),
$$
which we obtain after $i_3-i_2-1$ Weyl reflections. If, on the
contrary, $i_3-i_2 +1-{k_3 n \over s}<0$, we must start from the
weight: 
$$
(\cdots,1,{\buildrel i_1+1 \over 2},1,
\cdots,1,\buildrel i_2
\over {i_3-i_2 +1-{k_3 n \over s}},{k_3 n \over s}+1-i_2-i_3,1,
\cdots,1,{\buildrel i_3 \over 2},1, \cdots),
$$
and apply  more reflections to it. We have two possibilities:

1) $i_3-i_1 -1-{k_3 n \over s} \ge 0,$

2) $i_3-i_1 -1-{k_3 n \over s}<0.$

In the first possibility we obtain a null state. In the second one,
after $i_2-i_1-2$ reflections, we arrive at:
$$
( \cdots,1,{\buildrel i_1+1 \over 2},i_3-i_1 -1-{k_3 n \over
s},{k_3 n \over s}+2-i_3+i_1,1, \cdots,1,{\buildrel i_2+1
\over 2},1, \cdots ).
$$
Now applying $\sigma_{i_1 +2}$ we obtain the following weight:
$$
( \cdots,1,\buildrel i_1+1 \over {i_3-i_1 -{k_3 n \over
s}+1},{k_3 n \over s}+1-i_3+i_1,1, \cdots,1,{\buildrel
i_2+1 \over 2},1, \cdots).
$$
Hence, if $i_3-i_1 -{k_3 n \over s}+1>0$, this component is positive
and our task is finished. Notice that we have supposed  $i_3-i_1
-1-{k_3 n \over s}<0$. Therefore,
$$ 
i_3=i_1 +{k_3 n \over s}.
$$
If, on the contrary, $i_3-i_1 -{k_3 n \over s}+1<0$, we need more reflections.
Again there are two subcases:

2.1) $i_3-{k_3 n \over s} \ge 0$,

2.2) $i_3-{k_3 n \over s}<0$.

In the first subcase we arrive at a null state. In the second
subcase, after
$i_1-2$ Weyl reflections, we obtain the weight:
$$
(i_2-i_1+{k_1-k_2 \over s}n,{k_2 n \over s}+1-i_2,i_3-{k_3 n \over s},{k_3 n
\over s}-i_3+2,1, \cdots,1,{\buildrel i_1+2 \over 2},1,
\cdots,1,{\buildrel i_2+1 \over 2},1, \cdots ),
$$ 
which, after applying $\sigma_3$, becomes:
$$
(i_2-i_1+{k_1-k_2 \over s}n, i_3-i_2+{k_2-k_3 \over s}n,{k_3 n \over
s}+1-i_3,1,\cdots,1,{\buildrel i_1+2 \over 2},1, \cdots).
$$ 
Again, we must arrange the components through Weyl reflections so
that all of them become positive. Notice that every
arrangement obtained in this way can be associated to a different
ordering. We have, for example, the following equivalences: 
$$ 
\eqalign{
i_2-i_1+{k_1-k_2 \over s}n \rightarrow i_2 \succ i_1,
\cr i_3-i_2+{k_2-k_3 \over s}n \rightarrow i_3 \succ i_2. \cr}
$$
In this way, every Weyl reflection we do for changing the sign of a component
can be understood as a permutation of this ordering. 
If we act on the preceding weight with the reflection $\sigma_1$
we obtain: 
$$
(i_1-i_2+{k_2-k_1 \over s}n, i_3-i_1+{k_1-k_3 \over s}n,{k_3 n \over s}+1-i_3,
\cdots ),
$$
corresponding to the ordering:
$$
i_3 \succ i_1 \succ i_2.
$$
Now recall that  another possibility for the second index remains, $i_2 = i_1
+{k_2 n \over s}$. The weight we start with is:
$$
(\cdots,1,{\buildrel i_1 \over 1},1, \cdots,1,{\buildrel i_2
\over 2},1, \cdots,1,1-{k_3 n \over s},1+{k_3 n \over s},1,
\cdots).
$$
Again, we follow the same steps, and we have the possibility
$i_3=i_2 +{k_3 n \over s}$, which does not lead to a null state. 
In this case we arrive at the weight:
$$
({k_1 n \over s}+1-i_1,1, \cdots,1,{\buildrel
i_1 \over 1},1,\cdots,1,{\buildrel i_2 \over 1},1,
\cdots,1,{\buildrel i_3 \over 2},1, \cdots),
$$
after $i_3-i_2-1$ Weyl reflections. In the case $i_3-i_2 +1-{k_3 n \over
s}<0$ we can sum $i_2-2$ units to this quantity, so we have two possibilities:

1) $i_3-1-{k_3 n \over s} \ge 0$,
     
2) $i_3-1-{k_3 n \over s}<0$.

In the first case we find a null state. In the second one we obtain the weight:
$$
({k_1 n \over s}+1-i_1, i_3-1-{k_3 n \over s},{k_3 n \over s}+2-i_3,
\cdots,1,{\buildrel i_2+1 \over 2},1, \cdots),
$$
which, after  applying $\sigma_2$, leads to:
$$
(i_3-i_1+{k_1-k_3 \over s}n,{k_3 n \over s}+1-i_3,1, \cdots,1,{\buildrel
i_2+1 \over 2},1, \cdots).
$$
Of course, we must rearrange this weight to have positive components
by performing new Weyl reflections. But the most important thing to note here
is that when
$i_2=i_1 +{k_2 n \over s}$, one does not have the possibility of setting 
$i_3=i_1 +{k_3 n \over s}$ in order to obtain a non null state. It is easy to
realize that for the rest of the indices one finds  the same behavior. Also
notice that in the last case we needed $i_3-2$ Weyl reflections to obtain the
final form of the weight. We would have expected $i_3-3$, but the fact that
$i_2$ is a type II index adds one more. 

Let us briefly analyze the general case for the Weyl reflections associated to
type I indices. Consider the following weight resulting from our reduction
procedure applied to the first $r$ indices. These consist of $r-k$  type I
indices and $k$ type II ones:
$$
\eqalign{
& \rho +(i_{p_2}-i_{p_1}+{k_{p_1}-k_{p_2} \over s} n) \lambda_1+ \cdots +
({k_{p_{r-k}} \over s}n-i_{p_{r-k}})\lambda_{r-k}+ \cr
& + \lambda_{i_{\mu_1}+r-k-1}+
\cdots +\lambda_{i_{\mu_{r-k}}}-k_{r+1}\lambda_{i_{r+1}-1}+
k_{r+1}\lambda_{i_{r+1}}+ \cdots \cr} 
$$ 
Suppose that the $r$-th index is type I, so $i_{r+1}=i_{p_{r-k+1}}$.
Taking  the presence of the $k$ weights
$\lambda_{i_{\mu_1}+r-k-1},\cdots,\lambda_{i_{\mu_k}-k}$ into account, we must
perform
$r-k+1$ sequences of Weyl reflections. The numbers of Weyl reflections in each
sequence are the following:
$$
\eqalign{
&i_{r+1}-1-i_{\mu_{r-k}}, \cr
&i_{\mu_{r-k}}-(i_{\mu_{r-k-1}}+1), \cr
&(i_{\mu_{r-k-1}}+1)-(i_{\mu_{r-k-2}}+2),\cr
&\,\,\,\,\,\,\,\,\,\,\,\,\,\,\,\,\,\,\,\,\ \vdots \cr
&i_{\mu_2}+r-k-2-(i_{\mu_1}+r-k-1),\cr
&i_{\mu_1}+r-k-1-(r-k).\cr}
$$
The first sequence of reflections produces the coefficient:
$$
1-k_{r+1}+i_{r+1}-i_{\mu_{r-k}},
$$
for the fundamental weight $\lambda_{i_{\mu_{r-k}}}$. For the rest of the 
weights a similar coefficient is obtained from the rest of the sequences.
 Summing all these reflections together, we get the number:
$$
i_{r+1}-(r+1)+k.
$$
Taking  these facts into account, the number of Weyl reflections associated to a
type I index $i_{\lambda_p}$ (before the reordering of the first $r-k$
coefficients, which gives an extra contribution, as we will show later) is:
$$
i_{\lambda_p}-\lambda_p+\xi(\lambda_p),
$$
where $\xi(\lambda_p)$ is the number of type II index preceding $i_{\lambda_p}$.

We have seen that if we do not want to have a null state the
indices must be either  type I (\ie, $i_{\lambda} \le {k_{\lambda} n \over s}$)
or  type II ($i_{\lambda}= i_{\mu}+{k_{\lambda} n \over s}$, $\mu<\lambda$). In
addition, if one index has taken a type II value, no next index can use the
same $i_{\mu}$. So we have a tree diagram like this: 
$$
 i_1 \le {k_1n \over s} \cases{ i_2
=i_1+{k_2n \over s} {\cases { i_3 \le {k_3n \over s} \cr
 i_3=i_2+{k_3n \over s}\cr}}\cr
 i_2\le {k_2n \over s} {\cases { i_3=i_2+{k_3n \over s}\cr
                                i_3=i_1+{k_3n \over s}\cr
i_3 \le {k_3n \over s}\cr}}\cr}
$$
We have given necessary conditions to have a contribution. Now we obtain the
sufficient conditions and the canonical representative. Suppose we have a
partition with cardinal $r$. Hence, we have  $r$ indices, from which $r-k$ are
type I and $k$ type II. From the preceding analysis we see that the
canonical representative has the form:    $$ \rho +  p_1 \lambda_1 + p_2
\lambda_2 + \cdots
+p_{r-k}\lambda_{r-k}+\lambda_{i_{\mu_1} +r-k-1}
+ \lambda_{i_{\mu_2}+r-k-2}+ \cdots +\lambda_{i_{\mu_{r-k}}}.
\eqn\diotima
$$
We must give the values of $p_i$ and $\mu_i$. Let us carry it out in turn.

 1) Notice that the $p_i$ depend only on the type I indices, which we denote by
$i_{\lambda_1},i_{\lambda_2}, \cdots, i_{\lambda_{r-k}}$, ordered according to
their natural ordering. The arrangement that one first finds applying our
procedure is, $$  
(i_{\lambda_2}-i_{\lambda_1}+{k_{\lambda_1}-k_{\lambda_2} \over
s}n, i_{\lambda_3}-i_{\lambda_2}+{k_{\lambda_2}-k_{\lambda_3} \over s}n,
\cdots,i_{\lambda_{p +1}}-i_{\lambda_p}+{k_{\lambda_p}-k_{\lambda_{p+1}}
\over s}n, \cdots ).
\eqn\sylvie
$$
If for some pair of indices $i_{\lambda_p}$,
$i_{\lambda_q}$, one has $i_{\lambda_p}
-i_{\lambda_q}+{k_{\lambda_q}-k_{\lambda_p} \over s}n =0$, then one can always
apply Weyl reflections and put them together, obtaining in this way a zero
component. This gives us the sufficient condition we stated in the theorem. If
this is not the case, one must find the index arrangement giving positive
components. The order relation is obviously: $$
 i_{\lambda_p} \succ i_{\lambda_q}\,\ {\rm iff} \,\
i_{\lambda_p}-i_{\lambda_q}+{k_{\lambda_q}-k_{\lambda_p} \over
s}n>0, 
\eqn\caterina
$$
so if we introduce the permutation $\tau$: 
$$ 
\tau = \left(
\matrix{i_{\lambda_1}&i_{\lambda_2}&\cdots&i_{\lambda_{r-k}}\cr
i_{\tau(\lambda_1)}&i_{\tau(\lambda_2)}&\cdots&i_{\tau(\lambda_{r-k})}
\cr}\right), \eqn\alsacia
$$
we obtain the values of $p_i$ we
have given in \laspes. Recall that in this structure we have an equivalence
between Weyl reflections and permutations; in fact, a Weyl reflection acts as a
transposition of two indices. So the sign associated to these additional Weyl
reflections is just the signature of $\tau$.
 
2) Now let us determine the $\mu_i$. From what we have seen, neither one of
them can be one of the $k$ quantities $\mu$ such that some
$i_{\lambda}$ verifies $i_{\lambda}=i_{\mu}+{k_{\lambda} n \over s}$. In fact,
the $\mu_i$ are the subindices of the remaining $r-k$ indices, ordered in
the natural way. We can prove this assertion inductively. Obviously this is
true for the first index, because in that case $\mu_1 = i_1$. Suppose  we have
followed our procedure to obtain positive components with the first $r$, from
which $k$ take type II values. We have the following weight, 
$$
\eqalign{
\rho + \cdots
+\lambda_{i_{\mu_1} +r-k-1} + \cdots +&
\lambda_{i_{\mu_{p-1}}+r-k-p+1} \cr
& +\lambda_{i_{\mu_p} +r-k-p}
+\lambda_{i_{\mu_{p+1}}+r-k-p-1}+ \cdots +\lambda_{i_{\mu_{r-k}}} + \cdots \cr}
$$
Now, if we do the same thing with the next component, corresponding to the
index $i_{r+1}$, we have two possibilities: either this is an index of type I,
or an index of type II with the form $i_{r+1}=i_{{\mu_p}+r-p} + {k_{r+1} \over
s}n$. In the first case we have: 
$$
\eqalign{
 \rho + \cdots +\lambda_{i_{\mu_1} +r-k} &
+ \cdots + 
\lambda_{i_{\mu_{p-1}}+r-k-p+2}  \cr 
& +\lambda_{i_{\mu_p} +r-k-p+1} +\lambda_{i_{\mu_{p+1}}+r-p}+ \cdots
+\lambda_{i_{\mu_{r-k}}+1} + \lambda_{i_{r+1}} + \cdots \cr}
$$
where $i_{r+1}>i_{\mu_{r-k}}$, so $\mu_{r-k+1}=r+1$ as we claim. In the second
case, after the necessary Weyl reflections, we obtain: $$  
 \rho + \cdots +\lambda_{i_{\mu_1} +r-k-1}
+ \cdots +
\lambda_{i_{\mu_{p-1}}+r-k-p+1}+ \lambda_{i_{\mu_{p+1}}+r-k-p}+ \cdots
+\lambda_{i_{\mu_{r-k}}+1} + \lambda_{i_{r+1}} + \cdots 
$$
where $\hat \mu_1=\mu_1$, $\cdots$, $\hat
\mu_{p-1}=\mu_{p-1}$, $\hat\mu_p = \mu_{p+1}$, $\cdots$, $\hat
\mu_{r-k-1}=\mu_{r-k}$, $\hat \mu_{r-k}= r+1$, which means that,  again,
 the indices are ordered as
stated in the theorem: $i_{\mu_1}< \cdots<i_{\mu_{p-1}}<i_{\mu_{p+1}}<
\cdots<i_{\mu_{r-k}}<i_{r+1}$.

 It remains to give the sign inherited from the Weyl reflections. The type I
indices give two contributions. The first one comes from the
$i_{\lambda_p}-\lambda_p + \xi(\lambda_p) $ Weyl reflections associated to each 
$i_{\lambda_p}$. The second one comes from the signature of the permutation
$\tau$ \alsacia. For  type II indices we also find two contributions.
Let these $k$ indices be $i_{\nu_p}$, $p=1, \cdots,k$, and consider the $k$
indices $i_{\hat \nu_p}$ such that $i_{\nu_p}=i_{\hat \nu_p}+{k_{\nu_p} n \over
s}$. If we fix both it is easy to see that we can choose different
correspondences between them, producing of course different weights. Each
correspondence can be understood as a permutation of the indices $i_{\hat
\nu_p}$. This goes as follows: we arrange the $i_{\nu_p}$ in the natural
ordering, and we associate  the corresponding $i_{\hat \nu_p}$ to them:
$$ 
\eqalign{ &i_{\nu_1}<i_{\nu_2}< \cdots < i_{\nu_k} \cr &i_{\hat
\nu_1}\,\,\,\,\,\ i_{\hat \nu_2} \,\,\,\,\,\ \cdots \,\,\,\,\ i_{\hat
\nu_k}\cr} 
$$
If we arrange the indices $i_{\hat \nu_p}$ in the natural way:
$i_{\sigma^{-1}({\hat\nu_1})}<i_{\sigma^{-1}({\hat\nu_2})}<\cdots
<i_{\sigma^{-1}({\hat\nu_k})}$, we obtain the  permutation:
$$
\sigma=\left( \matrix
{i_{\sigma^{-1}({\hat\nu_1})}&i_{\sigma^{-1}({\hat\nu_2})}& \cdots
&i_{\sigma^{-1}({\hat\nu_k})}\cr
                         i_{\hat \nu_1}& i_{\hat \nu_2}&\cdots &
i_{\hat \nu_k}\cr}\right).
$$    
Now let $i_r$ be the first index taking a type II value. Before applying the
chain of reflections $\sigma_{i_{r-1}}, \cdots $ we have the following weight,
resulting from previous reflections: 
$$
 \rho + \cdots+ \lambda_{i_1 +r-2}
+ \lambda_{i_2+r-3}+\lambda_{i_{p-1}+r-p}+\lambda_{i_p
+r-p-1} +\lambda_{i_{p+1}+r-p-2} \cdots +\lambda_{i_{r-1}}.
$$  
Suppose $i_r=i_p+{k_r n \over s}$. The referred chain gives:
$$
 \rho + \cdots+ \lambda_{i_1 +r-2}
+ \lambda_{i_2+r-3}+\lambda_{i_{p-1}+r-p}+ +\lambda_{i_{p+1}+r-p-1}
\cdots +\lambda_{i_{r-1}+1} + \lambda_{i_r},
 $$  
obtained after $i_r-1-(i_p+r-p-1)=i_r-i_p+p-r$ Weyl reflections. Now let 
$i_t$ be the second type II index, verifying $i_t=i_q+{k_t n \over s}$.
Recalling the previous discussion, we have two possibilities: either $ i_q>i_p$
or $i_q<i_p$. In the first case, which keeps the natural ordering, the index 
$i_q +r-q$  shifts to $i_q +r-q +t-1-r =i_q+t-1-q$, and in the following chain
we need $i_t-1-(i_q +t-1-q)=i_t-i_q+q-t$ Weyl reflections. In the other
case, on the contrary, the shift gives $r-q-1+t-1-r=t-2-q$ and we need one more
Weyl reflection: $i_t-i_q+q-t+1$. This corresponds to the transposition: 
$$ 
(i_p \,\ i_q) \rightarrow (i_q \,\ i_p). 
$$
Therefore, we have two contributions to the total number of Weyl reflections.
The first  is the only one appearing when the permutation is the identity: 
$$
\sum_{p=1}^k  i_{\nu_p}-i_{\hat \nu_p} + \hat \nu_p - \nu_p \eqn\albertine,
$$
where the sum runs over the type II indices. But for a non trivial
permutation we have to take into account the transpositions of the form $(i \,\
i+1)$ into which it decomposes. Each of them corresponds to an additional Weyl
reflection. For example, the permutation $$ \left( \matrix {1&2&3\cr
                 3&2&1\cr}  \right),
$$       
gives three Weyl reflections, corresponding to the transpositions:
$$
(123) \rightarrow (213) \rightarrow (231),
$$ 
for the index $i_{\hat \nu_1}$, and to the single transposition:
$$
(231) \rightarrow (321),
$$
for the index $i_{\hat \nu_2}$. The additional sign is simply given by the
signature of the permutation  $\sigma$. 

 Putting all this together, we find the sign associated to the canonical
representative  \diotima:
$$
\epsilon (\tau) \epsilon (\sigma) (-1)^{\sum i_{\lambda_p}-\lambda_p +
\xi(\lambda_p) + \sum i_{\nu_p}-i_{\hat \nu_p} +  \hat \nu_p - \nu_p},
$$
where the first sum runs over the type I indices and the second one over the
type II indices. We can put it together as:
$$
\sum_{p=1}^{r-k} i_{\mu_p}-\mu_p + \xi(\lambda_p),
$$
where the sum runs over the set of indices {\it not}
verifying $i_{\lambda}=i_{\mu}+{k_{\lambda} n \over s}$ (the same which appears
in \diotima). This completes the proof of the theorem.

Now we state and prove the main result of this paper.

{\bf Theorem 4.2.} The HOMFLY polynomial for a torus link $(n,m)$ with $s$
components is given by:
$$ 
\eqalign{ 
&X^{(n,m)}={\lambda^{{1 \over 2}(n-1)(m-1)} (t-1) 
\over \lambda t -1} \sum_{(k_{\lambda})}{s! \over \prod_{\lambda}^r
k_{\lambda}!}t^{{mn \over 2}
(1-{1 \over s^2}\sum_{\lambda=1}^r
k_{\lambda}^2)}\cr
&\times 
\sum_{{\cal I}_{(k_\lambda)}}\epsilon (\tau) \epsilon  (\sigma)
(-1)^{\sum_{p=1}^{r-k} i_{\mu_p}-\mu_p+\xi(\lambda_p)} t^{{ m \over s}
(\sum_{\lambda=1}^r k_{\lambda}i_{\lambda} -s)}\cr
& \times \lambda^{ n\over
2}\prod_{\tau=1}^{r-k-1}[p_{\tau}+1] \cdots [\sum
_{\lambda=\tau}^{r-k-1} p_{\lambda}+r-k-1] 
 \prod_{1 \le \sigma<\tau \le
r-k} [i_{\mu_{\tau}}-i_{\mu_{\sigma}}]\cr
& \times \prod_{\tau=1}^{r-k} {
[i_{\mu_{\tau}}] \over \prod_{j=1}^{r-k}[\sum_{\lambda=\tau}^{r-k}
p_{\lambda}+i_{\mu_j}+r-k-\tau] } \Bigg[ {N+ \sum_{\lambda=\tau}^{r-k}
p_{\lambda}+r-k-\tau \atop \sum_{\lambda=\tau}^{r-k}
p_{\lambda}+r-k-\tau} \Bigg] \Bigg[ {N \atop i_{\mu_{\tau}} } \Bigg].
\cr} \eqn\cocotier
$$ 
    In this formula the first sum runs over the ordered partitions of $s$ 
which are denoted by $(k_\lambda)$. For each partition $(k_\lambda)$ its
cardinal is labeled by $r$ (this $r$ should not be confused with the rank of 
the gauge group as used in Appendix A). 
The second sum, subordinate to the previous one, runs
over the set of arrangements of indices ${\cal I}_{(k_\lambda)}$ constructed 
in the previous theorem. Each arrangement 
of indices in this set, denoted by $\mu_i$, $i=1,\cdots,r$,
has $k$ indices of type II, as  specified by the previous theorem. 
For each arrangement of indices, the permutations $\tau$ and $\sigma$, 
whose signatures  appear in \cocotier, as well as the quantities 
$\xi(\lambda_p)$, 
$p=1,\cdots,r-k$, are given  in parts 1), 2) and 3) 
of the previous theorem, respectively. Similarly,  the quantities 
$p_i$, $i=1,\cdots,r-k$, are also given in that theorem by equation \laspes.
Notice that this expression only depends on the
 variables $t$ and $\lambda$ because, as shown in \karina,
the $q$-combinatorial numbers that appear in \cocotier\ can be
reexpressed in such a way that $N$ is hidden.

{\bf Proof.} The proof of this theorem goes along the same lines as
 that of theorem 3.1. Using the same arguments that led to \eloisa, 
we must first compute the character of a weight of the form: 
$$
\Lambda=\rho +  p_1 \lambda_1 + p_2 \lambda_2 +
\cdots
+p_{r}\lambda_{r}+\lambda_{i_1 +r-1}
+ \lambda_{i_2+r-2}+ \cdots +\lambda_{i_r},
$$
which corresponds to an arrangement of indices in ${\cal I}_{(k_\lambda)}$ 
as shown in the previous theorem.
We use again the expression \luiza, so it is necessary to classify the 
positive roots according to their contribution to the product. 

1) First we consider positive roots beginning with $\alpha_i$,
$1 \le i < r-1$ and ending with $\alpha_j$, $i<j \le r-1$. They have the
structure: $$
\eqalign{
&\alpha_1,\cr
&\alpha_1 +\alpha_2, \cr
&\,\,\ \vdots\cr
&\alpha_1 + \cdots + \alpha_{r-1},\cr
&\alpha_2, \cr
&\alpha_2+\alpha_3,\cr
&\,\,\ \vdots\cr
&\alpha_2+ \cdots +\alpha_{r-1},\cr
&\,\,\ \vdots\cr
&\alpha_{r-1}.\cr}
$$
 The contribution of the group beginning with $\alpha_k$ is:
$$
{[p_k+1][p_k+p_{k+1}+2] \cdots [p_k+
\cdots+p_{r-1}+r-k] \over [1][2] \cdots [r-k]} 
$$
Putting them together, with $1 \le k \le r-1$, we obtain:
$$ 
{\prod_{k=1}^{r-1}[p_k+1][p_k+p_{k+1}+2] \cdots
[\sum_{\lambda=k}^{r-1} p_{\lambda}+r-1] \over \prod _{k=1}^{r-1}
[r-k]!}.
$$

 2) The next group consists of the positive roots beginning with
$\alpha_1, \cdots, \alpha_r$ and ending with $\alpha_{i_1+r-2},
\cdots, \alpha_{i_r-1}, \cdots, \alpha_{N-1}$. The contribution of
those beginning in $\alpha_1$ is: 
$$
\eqalign{
\alpha_1+\cdots+\alpha_r+\cdots+  \alpha_{i_1+r-2}  :{\hbox{\hskip10pt}}
&{[\sum_{\lambda=1}^r p_{\lambda}+i_1+r-2]! \over[\sum_{\lambda=1}^r
p_{\lambda}+r-1]!} {[r-1]! \over [i_1+r-2]!},\cr
\alpha_1+\cdots+\alpha_{i_1+r-1} +\cdots+ \alpha_{i_2+r-3} 
:{\hbox{\hskip10pt}} &{[\sum_{\lambda=1}^r p_{\lambda}+1+i_2+r-3]! \over
[\sum_{\lambda=1}^r p_{\lambda}+1+i_1+r-2]!} {[i_1+r-2]! \over[i_2+r-3]!},\cr
&\,\,\ \vdots\cr
\alpha_1+\cdots+\alpha_{i_r} +\cdots+ \alpha_{N-1}  : {\hbox{\hskip10pt}}
 &{[\sum_{\lambda=1}^r
 p_{\lambda}+r+N-1]! \over [\sum_{\lambda=1}^r
p_{\lambda}+r+i_r-1]!}{[i_r-1]! \over [N-1]!}, \cr}
$$
where we have explicitly quoted the positive roots corresponding to each
contribution. Multiplying all these factors we obtain:  $$
{[\sum_{\lambda=1}^r p_{\lambda}+r+N-1]! \over [\sum_{\lambda=1}^r 
p_{\lambda}+i_1+r-1] \cdots [\sum_{\lambda=1}^r p_{\lambda}+i_r+r-1]
[\sum_{\lambda=1}^r p_{\lambda}+r-1]!}{[r-1]! \over [N-1]!}.
$$
The contributions from the positive roots beginning with $\alpha_2,
\cdots, \alpha_k, \cdots, \alpha_r$ are very similar, with the only difference
that the sum of the $p_{\lambda}$ now begins with $\lambda=k$, $2 \le k \le r$,
and that there is a shift of $k$ units in the products. Hence the contribution
from the positive roots beginning with $\alpha_k$ is: 
$$ {[\sum_{\lambda=k}^r
p_{\lambda}+r+N-k]! \over [[\sum_{\lambda=k}^r  p_{\lambda}+i_1+r-k] \cdots
[\sum_{\lambda=1}^r p_{\lambda}+i_r+r-k] [\sum_{\lambda=k}^r
p_{\lambda}+r-k]!}{[r-k]! \over [N-k]!}. 
$$

 3) The last group of positive roots consists of the positive roots beginning
with $\alpha_{r+1}$ and ending with $ \alpha_{i_2+r-3}, \cdots, \alpha_{N-1}$,
those beginning with $\alpha_{i_1+r}$ and ending with $\alpha_{i_3+r-4},
\cdots, \alpha_{N-1}$, etc. We find here two kinds of contributions:
 
3.1) Those from the positive roots ending with $\alpha_{N-1}$. They are:
$$
\eqalign{
\alpha_{r+1} +\cdots+\alpha_{N-1}: &
\prod_{\mu=i_r+1}^N \prod_{\lambda=r+1}^{i_1+r-1}
{[\mu-\lambda+r] \over [\mu-\lambda]}, \cr
&\,\,\ \vdots\cr
 \alpha_{i_{r-1}+2} +\cdots+\alpha_{N-1} 
:& \prod _{\mu=i_r+1}^N
\prod_{\lambda=i_{r-1}+2}^{i_r}{[\mu-\lambda+r]
 \over [\mu-\lambda]},\cr}
$$
Note that, if $N$ is great enough, we can make use of the following equality:
$$
\eqalign{
& \prod_{\mu=p+1}^q \prod_{\lambda=r }^s
{[\mu-\lambda+m] \over [\mu-\lambda]}= \cr
&= \prod_{\lambda=r}^s {[q-\lambda+m] \over [q-\lambda]} \cdots
{[q-\lambda] \over [q-\lambda-m]} \cdots {[p+2m-\lambda] \over [p+m-
\lambda]} \cdots {[p+1-\lambda+m] \over [p+1- \lambda]} \cr
&=\prod_{\lambda=r}^s {[q-\lambda+m] \over
[p+m-\lambda]}{[q-1+m-\lambda] \over [p-1+m-\lambda]} \cdots
{[q-\lambda] \over [p+1-\lambda]}.\cr}
$$
In this case one obtains:
$$
\eqalign{
\alpha_{r+1}+\cdots+\alpha_{N-1}  :
&\prod_{\lambda=r+1}^{i_1+r-1} {[N-\lambda+r] \over
[i_r+r-\lambda]}{[N-\lambda+r-1] \over [i_r-\lambda+r-1]} \cdots
{[N-\lambda+1] \over [i_r+1-\lambda]}, \cr
&\,\,\ \vdots\cr
\alpha_{i_k+r-k+1} +\cdots+ \alpha_{N-1}  :
&\prod_{\lambda=i_k+r-k+1}^{i_{k+1}+r-k-1} {[N-\lambda+r-k] \over
[i_r+r-\lambda-k]} \cdots {[N-\lambda+1] \over [i_r+1-\lambda]}, \cr
&\,\,\ \vdots\cr
\alpha_{i_{r-1}+1} +\cdots+ \alpha_{N-1}  :
&\prod_{\lambda=i_{r-1}+1}^{i_r} {[N-\lambda+1] \over
[i_r+1-\lambda]}.\cr} $$
Putting all these products together we arrive at:
$$
\prod_{\lambda=r+1}^{i_r} {[N-\lambda+1] \over [i_r+1-\lambda]} \cdots
\prod_{\lambda=r+1}^{i_{r-k+1}+k-1} {[N-\lambda+k] \over
[i_r+\lambda+k]} \cdots \prod_{\lambda =r+1}^{i_1+r-1} {[N-\lambda+r]
\over [i_r-\lambda+r]}.
$$

 3.2) We have also the positive roots not ending with
$\alpha_{N-1}$. We have first those beginning with 
$\alpha_{r+1}$ and ending with $\alpha_{i_k +r-k}$, $1 \le k \le r$.
This contribution has the form:
$$
\eqalign{
&\prod_{1 \le k <r} \prod_{\mu=i_k+r-k+1}^
{i_{k+1}+r-k+1}\prod_{\lambda=r+1 }^{i_1+r-1} {
[\mu-\lambda+k] \over [\mu-\lambda]}= \cr &=\prod_{1 \le k <r}
\prod_{\mu=i_k+r-k+1}^ {i_{k+1}+r-k+1}{[\mu-(r+1)+k]! \over
[\mu-(i_1+r-1)+k-1]!}{[\mu-(i_1+r-1)-1]! \over [\mu-(r+1)]!}=\cr
&=\prod_{1 \le k <r} \prod_{\mu=i_k+r-k+1}^
{i_{k+1}+r-k+1} {[\mu-r-1+k][\mu-r-1+k-1] \cdots [\mu-r] \over
[\mu-i_1-r+k] [\mu-i_1-r+k-1] \cdots [\mu-i_1-r-1]}=\cr
&=\prod_{1 \le k<r}{[i_{k+1}-2]! \over [i_k-1]!} \cdots {[i_{k+1}-k-1]!
\over [i_k-k]!} \prod _{1 \le k<r} {[i_k-i_1]! \over [i_{k+1}-1-i_1]!}
\cdots {[i_k-i_1-k+1]! \over [i_k+1-k-i_1]!}= \cr
&={[i_r-2]! \cdots [i_r-r]! \over [i_1-1]! \cdots [i_{r-1}-r]!}
{\prod_{\lambda=2}^{r-1} [i_{\lambda}-i_1]! \over \prod_{\lambda=2}^r
[i_r-i_1-\lambda+1]!}. \cr}
\eqn\gilberte
$$
 Second, we have the contribution from the positive roots beginning with
$\alpha_{i_j+r-j}$ and ending with $\alpha_{i_k+r-k}$. Here we have:
$$
\prod_{1 \le j<k<r} \prod_{\lambda=i_j+r-j+1}^ {i_{j+1}+r-j-1}
\prod_{\mu=i_k+r-k+1}^
{i_{k+1}+r-k+1}{[\mu-\lambda+k-j] \over [\mu-\lambda]}. $$
We can arrange this in the same way as in the previous case:
$$
\eqalign{
&\prod_{1 \le j<k<r}{[i_{k+1}-i_j-2]! \over [i_k-i_j-1]!} \cdots
{[i_{k+1}-i_j+j-k-1]! \over [i_k-i_j+j-k]!} \cr
& \times \prod_{1 \le j<k<r}
{[i_k-i_{j+1}]! \over [i_{k+1}-i_{j+1}-1]!}{[i_k-i_{j+1}+j-k+1]! \over
[i_{k+1}-i_{j+1}+j-k]!}, \cr }
$$
and, finally we obtain:
$$
\prod_{1 \le j \le r-2}{\prod_{\lambda=2}^{r-j} [i_r-i_j-\lambda]!
\over \prod_{\lambda=j+1}^{r+1}
[i_\lambda-i_j-1]!}{\prod_{\lambda=j+1}^{r-1}[i_\lambda-i_{j+1}]!
\over \prod_{\lambda=2}^{r-j}[i_r-i_{j+1}-\lambda+1]!}.   
$$
Now, we put this together with the second factor in \gilberte. After some
straightforward calculations we obtain: $$
{\prod_{1 \le j \le r-2}\prod_{ j<\lambda \le r-1}
[i_{\lambda}-i_j] \over \prod _{j=1}^{r-1}[i_r-i_j-1]!}.
\eqn\oriane
$$
 The contribution to the character from 1) and 2) can be written as:
$$
\eqalign{
&\prod_{k=1}^{r-1}[p_k+1] \cdots [\sum _{\lambda=k}^{r-1}
p_{\lambda}+r-1]\cr
 \times & \prod_{k=1}^r {[\sum_{\lambda=k}^r
p_{\lambda}+r+N-k]! [i_r-i_k]! \over
\prod_{\mu=1}^r[\sum_{\lambda=k}^r p_{\lambda}+i_{\mu}+r-k]
[\sum_{\lambda=k}^r p_{\lambda}+r-k]![N-i_k]![i_r-k]!}. \cr }
$$
 Now if we join this  with the first factor in \gilberte,
we arrive to:
$$
\eqalign{
&\prod_{k=1}^r {1 \over [N-i_k]![i_r-k]!}{[i_r-2]! \cdots [i_r-r]! \over
[i_1-1]! \cdots [i_{r-1}-r]!}=\cr
&=\prod_{k=1}^r [N]! [i_k] \Bigg[ {N \atop  i_k} \Bigg ].\cr}
$$
Introducing the $q$-combinatorial number: 
$$
{[\sum_{\lambda=k}^r p_{\lambda}+r+N-k]!  \over [\sum_{\lambda=k}^r
p_{\lambda}+r-k]!} = [N]!\Bigg[ {N+ \sum_{\lambda=k}^r
p_{\lambda}+r-k \atop \sum_{\lambda=k}^r p_{\lambda}+r-k} \Bigg ],
$$
we obtain, taking into account the remaining factors, the following expression
for the character: $$
\eqalign{
{\rm ch}_{\Lambda}=&\prod_{k=1}^{r-1}[p_k+1] \cdots [\sum _{\lambda=k}^{r-1}
p_{\lambda}+r-1] \prod_{1 \le j<k \le r} [i_k-i_j]\cr
& \times \prod_{k=1}^r {[i_k] \over \prod_{\mu=1}^r[\sum_{\lambda=k}^r
p_{\lambda}+i_{\mu}+r-k]}\Bigg[ {N+ \sum_{\lambda=k}^r p_{\lambda}+r-k
\atop {\sum_{\lambda=k}^r p_{\lambda}+r-k}} \Bigg] \Bigg[{ N \atop
i_k} \Bigg].\cr} 
\eqn\angela$$
This formula generalizes the one obtained in the previous section for
weights with the form $p\lambda_1 +\lambda_{i+1}$. 

It remains only to compute the phase factor appearing in the action of the knot
operator \dora, as well as the deframing factor in \equisdos. For an 
arrangement of indices in ${\cal I}_{(k_\lambda)}$ one finds,
$$ 
\eqalign{
\bigg( \sum \mu_i \bigg)^2 =& \bigg(\sum_{\lambda=1}^r k_{\lambda}
\mu_{i_{\lambda}}\bigg)^2 = \bigg( \sum_{\lambda=1}^r
k_{\lambda}^2\bigg ) \bigg(1-{1 \over N} \bigg)-{2 \over N} \sum
_{\lambda,\mu =1}^r k_{\lambda} k_{\mu} = \cr &=\sum_{\lambda=1}^r
k_{\lambda}^2 -{s^2 \over N},\cr} 
$$
where we have used (A20). In the same way, using now (A21), we obtain:
$$
\rho \cdot \bigg(\sum_{\lambda=1}^r k_{\lambda} \mu_{i_{\lambda}} \bigg)={1
\over 2} \sum_{\lambda=1}^r k_{\lambda}(N-2i_{\lambda}+1),
$$
and therefore the phase factor is:
$$
\exp \Bigg[ i \pi {nm \over s^2(k+N)} \bigg(\sum_{\lambda=1}^r
k_{\lambda}^2 -{s^2 \over N} \bigg) +2 \pi i { m \over s(k+N)}{1 \over
2} \sum_{\lambda=1}^r k_{\lambda}(N-2i_{\lambda}+1) \Bigg].
\eqn\phasefactor
$$ 
The deframing phase is the same one as in \pera. Taking  \angela, 
\phasefactor, \sigsig\ and \pera\ into account, and computing \equisdos, one
obtains immediately \cocotier. This ends the proof of theorem 4.2.

\endpage

\chapter{ The Jones and Alexander polynomials as the limits $N\rightarrow 
2$ and $N\rightarrow 0$}

It is well known [\jonesAM] that the Jones and the Alexander polynomials can be
recovered from the HOMFLY polynomial putting
$\lambda=t$ and $\lambda=t^{-1}$, respectively. In the Chern-Simons
approach with gauge group $SU(N)$ this corresponds to the limits $N\rightarrow
2$ and $N \rightarrow 0$, as \lalanda\ shows. In this section we analyse the
behaviour of the expression \cocotier\ in these two cases and compare the
results with the known ones [\jonesAM, \poli]. This is also a beautiful
exercise to  show how \cocotier\ works.

Let us begin with the  limit corresponding to the Jones polynomial. 
If we put $N=2$ in
\cocotier\ and we use the explicit expressions for the $q$-combinatorial
numbers we have given in
\karina, we easily see that the character becomes zero if $i_{\mu_{\tau}}
>2$. The only contributions come from the indices $i_{\mu_{\tau}}
\le 2$, so we only need partitions of cardinal two. The indices are:

1) partition $(s)$  $\rightarrow i_1=1, \,\ 2$,

2) partition $(l,s-l)$, $l \ge 1$ $\rightarrow i_1 =1, \,\ i_2 =2$.

\newdimen\tableauside\tableauside=1.0ex
\newdimen\tableaurule\tableaurule=0.4pt
\newdimen\tableaustep
\def\phantomhrule#1{\hbox{\vbox to0pt{\hrule height\tableaurule width#1\vss}}}
\def\phantomvrule#1{\vbox{\hbox to0pt{\vrule width\tableaurule height#1\hss}}}
\def\sqr{\vbox{%
  \phantomhrule\tableaustep
  \hbox{\phantomvrule\tableaustep\kern\tableaustep\phantomvrule\tableaustep}%
  \hbox{\vbox{\phantomhrule\tableauside}\kern-\tableaurule}}}
\def\squares#1{\hbox{\count0=#1\noindent\loop\sqr
  \advance\count0 by-1 \ifnum\count0>0\repeat}}
\def\tableau#1{\vcenter{\offinterlineskip
  \tableaustep=\tableauside\advance\tableaustep by-\tableaurule
  \kern\normallineskip\hbox
    {\kern\normallineskip\vbox
      {\gettableau#1 0 }%
     \kern\normallineskip\kern\tableaurule}%
  \kern\normallineskip\kern\tableaurule}}
\def\gettableau#1 {\ifnum#1=0\let\next=\null\else
  \squares{#1}\let\next=\gettableau\fi\next}

\tableauside=2.0ex
\tableaurule=0.4pt

In the second case we have two possibilities, because
$i_2=2$ can be either type I or type II. The second possibility only
occurs when $ {s-l \over s}n = 1$, \ie, $n=s$ and $s-l=1$. In case 1), the
corresponding weights are: 
$$
\eqalign{
(n-1)&\lambda_1 +\lambda_1=n\lambda_1,\cr
&(n-2) \lambda_1 + \lambda_2.\cr}
$$
In case 2) we have the weight:
$$
p_1 \lambda_1 +p_2 \lambda_2 + \lambda_2 + \lambda_2,
$$
if $i_2=2$ is  type I, and 
$$
(s-3)\lambda_1 + \lambda_2,
$$
if it is of type II. Notice that all the weights surviving when $N=2$
correspond to irreducible representations of $SU(2)$ since their highest
weight has only two non-vanishing components as in the last two expressions.
In terms of  Young tableaux, these representations have either only one
or two columns. 
For example, the weight $p_1 \lambda_1 +(p_2+2) \lambda_2$ corresponds to the
tableau: 
$$ 
\tableau{2 2 2 2 2 1 1 1}
$$
with $p_1+p_2+2$ squares in the first column and $p_2+2$ in the second one. 
This is in agreement  with the construction of the Jones
polynomial given in [\jonesAM]. The characters follow from the previous
formulae with $N=2$:
$$
\eqalign{ {\rm ch}_{n\lambda_1}= &t^{-{n \over 2}} {t^{n+1}
-1 \over t-1}, \cr
{\rm ch}_{(n-2) \lambda_1 + \lambda_2} =& t^{-{n
\over 2}} {t^n-t \over t-1}, \cr
{\rm ch}_{p_1 \lambda_1 +p_2 \lambda_2 + \lambda_2 + \lambda_2} =
&t^{-{p_1 \over 2}} {t^{p_1+1}-1 \over t-1}, \cr
{\rm ch}_{(s-3)\lambda_1 + \lambda_2}= &t^{-{s-1 \over 2}} {t^{s-1}-1
\over t-1}, \cr}
$$
where, according to theorem 4.1.:
$$
p_1=\cases { {2l-s \over s}n, &if $1+{ 2l-s  \over s}n >0$, \cr
             {s-2l \over s}n-2, &if $1+{2l-s \over s}n <0$. \cr }
\eqn\aitane 
$$
Notice that when $1+{2l-s \over s}n =0$ we have no contribution from the
weight. However, in this case the  corresponding character is
zero, so there is no need to modify the expressions. The factors in \cocotier\
which multiply the combinatorial numbers can be easily computed. For the
partition of cardinal $1$ one finds
$t^{m(i_1-1)}$, while, for the partitions of cardinal $2$, 
$$
t^{ {mnl(s-l) \over s^2} + {m \over s}(s-l)},
$$
is obtained.
We must also calculate  the signs produced by the Weyl reflections. For the
partition of cardinal $1$ the sign is $(-1)^{i_1-1}$, and for those having
cardinal $2$: 
$$ 
\epsilon(\tau)(-1)^{i_1-1+i_2-2}=\cases { 1, &if $1+{2l-s  \over
s}n >0$,\cr
             -1, &if $1+{2l-s  \over s}n <0$, \cr}
\eqn\dindin
$$
 when $i_2$ is  type I, and $(-1)^{i_2-2}=1$ in the other case. Notice that if
we set $n=s$, $l=s-1$, in the expressions corresponding to the type I index,
we obtain the values corresponding to type II. Hence, we can discuss both
cases together without further specification. Now if we calculate 
${\rm ch}_{p_1
\lambda_1 +p_2 \lambda_2 + \lambda_2 + \lambda_2}$ for the two possibilities
given in \aitane\ we obtain the same result  but with opposite signs: 
$$
\eqalign{ 1+{2l-s  \over s}n >0 \rightarrow  &t^{-{p_1\over 2}} (t^{p_1+1}-1) =
t^{{2l-s \over 2s}n+1}-t^{-{2l-s \over2s}n},\cr 1+{2l-s  \over s}n
<0\rightarrow  &t^{-{p_1\over 2}} (t^{p_1+1}-1)=t^{{s-2l \over 2s}n}-t^{-{s-2l
\over 2s}n+1}.\cr} 
$$  
Hence, taking into account the additional signs in \dindin, we
see that in both cases we obtain the same factor. Finally we substitute all
these results into \cocotier. The degeneracy factor appearing there is $1$ for
the partition of cardinal $1$ and a combinatorial number for the partitions of
cardinal $2$. We obtain, after some staightforward calculations:  $$
P(t)^{(n,m)}={t^{{1 \over 2}(n-1)(m-1)} 
\over 1-t^2} \sum_{l=0}^s{s \choose l}
t^{{m \over s }(1+{nl \over s})(s-l)} (t^{{n \over s}(s-l)} - t^{1+{n
\over s}l}), 
$$
which is just the expression given in [\poli].

Now we turn to the Alexander polynomial. This is a very particular limit
because it corresponds in some way to a fading of the gauge group. In fact, the
normalized vacuum expectation values $V^{(n.m)}_{\lambda_1}/ S_{\rho,\rho} $
and $V^{(1,0)}_{\lambda_1}/S_{\rho,\rho}$ become zero. However, if we take
their quotient we obtain a non-vanishing result which is the Alexander
polynomial for torus links. This is easily seen if we analyze the limit of the
$q$-combinatorial numbers when $N \rightarrow 0$ using \karina. 
Clearly:
$$ 
\Bigg[ {N +p \atop p} \Bigg] \rightarrow 1.
$$
The other quantity, $\Big[{ N \atop i} \Big]$, has always $\lambda t-1$ as
a factor, which goes to zero in the limit we are studying. However, the unknot
normalization contains an identical factor, so the quotient becomes: 
$$
{1 \over \lambda t-1}[i] \Bigg[ {N \atop i} \Bigg] \rightarrow
(-1)^{i-1}, \,\,\,\,\
{\hbox{\rm if}} \,\ N \rightarrow 0.
\eqn\amaya
$$
These facts imply that in the Alexander limit not all the weights appearing in
theorem 4.1. contribute. Each character having more than one
factor of the form $\Big[{ N \atop i} \Big]$ becomes null. There remain
only two kinds of weights: those corresponding to the partition with cardinal  
$1$, and those corresponding to partitions with cardinal $r$ with $k=r-1$, \ie,
with only one type I index. These weights are of the form: 
$$
({k_1n \over s}-i_1)\lambda_1 + \lambda _{i_r}, \,\,\,\,\ i_r=i_1+{n \over
s}\sum_{\lambda=2}^r k_{\lambda}.
$$
Notice that these weights correspond to irreducible representations of $SU(N)$
associated to Young tableaux with the following structure,
$$
\tableau{6 1 1 1 1 1}
$$
having ${k_1n \over s}-i_1+1$ boxes in the first column and $i_r$ boxes in
the first row. This agrees again with the results of [\jonesAM]. The sign from
the Weyl reflections for this kind of weights is given by $(-1)^{i_r-r}$.
Taking  \amaya\ into account we obtain the total sign $(-1)^{r+1}$. Hence,
the expression for tha Alexander polynomial is:  
$$
\eqalign{
 &A(t)^{(n,m)}={t^{-{1 \over 2}(n-1)(m-1)}(t-1) \over
t^n-1}\cr
&\times \sum_{(k_{\lambda})}{s! \over \prod_{\lambda=1}^r k_{\lambda}!}
\sum_{i_1=1}^{k_1n \over s} (-1)^{r+1}t^{{mn \over 2} (1-{1 \over
s^2}\sum_{\lambda=1}^r k_{\lambda}^2)+{ m \over s} (\sum_{\lambda=1}^r
k_{\lambda}i_{\lambda} -s)}.\cr}
\eqn\alexa 
$$ 
Let us compute the phase present in \alexa. One finds, 
$$
\eqalign{
\sum_{\lambda=1}^r k_{\lambda}i_{\lambda}=&k_1i_1+k_2(i_1+{k_2n \over
s})+ \cdots = i_1(k_1+ \cdots+k_r) +{n \over s}(k_2^2 + k_2k_3+
\cdots)=\cr
&=si_1 +{n \over s}(\sum_{\lambda=1}^r k_{\lambda}^2
+\sum_{\lambda,\mu} k_{\lambda}k_{\mu} -k_1s),\cr}
$$
and, adding the remaining factor, it becomes:
$$
mn +m(i_1-1)-{mnk_1 \over s}.
$$
Taking into account that,
$$
\sum_{i_1=1}^{k_1n \over s} t^{m(i_1-1)}= { 1-t^{mnk_1 \over s} \over
1-t^m},
$$
one finds, finally,
$$
A(t)^{(n,m)}={t^{-{1 \over 2}(n-1)(m-1)}t^{mn}(t-1) \over
(t^n-1)(t^m-1)}\sum_{(k_{\lambda})}{s! \over \prod_{\lambda=1}^r 
k_{\lambda}!}(-1)^{r+1}(t^{-{mnk_1 \over s}}-1).
$$
The sum over partitions present in this expression is computed in Appendix
B. The result is:
$$
\sum_{(k_{\lambda})}{s! \over \prod_{\lambda=1}^r 
k_{\lambda}!}(-1)^{r+1}(t^{-{mnk_1 \over s}}-1)=(t^{-{mn \over s}}-1)^s.
\eqn\pesa
$$ 
In this way we obtain the Alexander polynomial for torus links: 
$$
A(t)^{(n,m)}=t^{-{1 \over 2}(n-1)(m-1)}{(t-1)(1-t^{ mn \over s})^s \over
(t^n-1)(1-t^m)}.
\eqn\dilla
$$
This expression agrees with the one presented in  \REF\roz{L. Rozansky and H.
Saleur \journal\np&B389(93)365} [\roz] up to a normalization factor.
\endpage

\chapter{Final comments and remarks}

The formula \cocotier\ for the HOMFLY polynomial that we have obtained is
rather intricate. It involves a sum over the ordered partitions of $s$, the
number of components of the link, and for each partition one must build the set 
${\cal I}_{(k_\lambda)}$ as prescribed by theorem 4.1. In practice its
calculation is very laborious and it is much more convenient to implement it as
a computer routine using a symbolic language. This implementation has been
carried out and the corresponding computer routine is presented in Appendix C.

Chern-Simons gauge theory guarantees that the invariant 
$X^{(n,m)}$ in \cocotier\ corresponding to the HOMFLY polynomial is symmetric,
\ie, $X^{(n,m)}=X^{(m,n)}$. To verify this property using \cocotier\ is a highly
non-trivial problem. Using the computer routine in
Appendix C we have checked this property in many cases. The formula \cocotier\
must also present properties under a reversal of the orientation
[\jonesAM]. For torus links, this property becomes:
$$
X^{(n,m)}(\lambda,t) = X^{(n,-m)}({\lambda}^{-1},t^{-1}).
\eqn\reversal
$$
Again, it is highly non-trivial to verify this property explicitly making use of
\cocotier. It follows, however, from general arguments in Chern-Simons gauge
theory and therefore it holds for \cocotier.

The general formula \cocotier\ can be simplified enormously by considering
particular cases. For the case in which $n$ and $m$ are coprime integers one
has a single component, $s=1$, and it reduces to \nuria.
The case in which $m=0$ corresponds to the $n$-component unlink. Our formula
becomes the well known result\foot{Notice that we use the same notation as in
[\witCS] and therefore the replacements $t^{1\over 2}\rightarrow -t^{1\over 2}$
and $\lambda^{1\over 2}\rightarrow -\lambda^{1\over 2}$ have to be made before
comparing with results in [\jonesAM]}: 
$$
X^{(n,0)}=\Big({\lambda t - 1 \over \lambda^{1\over 2} (t-1) }\Big)^{n-1}.
\eqn\elenecero
$$

For links of the form, $(2,2p)$,
where $p$ is any integer, one has $s=2$, and \cocotier\ becomes:
$$
X^{(2,2p)}={\lambda^{2p-1\over 2}\over (t-1)}
{\lambda(t^2+t^{1+2p})-1-t^{1+2p}\over (t+1)}.
\eqn\casodosdosp
$$
For $p=1$ this link corresponds to the Hopf link and indeed, one finds 
the well known
result:
$$
X^{(2,2)}={\lambda^{1\over 2}\over (t-1)}(t^2\lambda - t^2 + t -1).
\eqn\hopflink
$$

Similar particular expressions can be obtained from \cocotier\ for other cases.
In Appendix D we present the result of evaluation of \cocotier\ for some
specific values of $n$ and $m$. From that table and from many other
calculations that we have performed one is led to conjecture that for any torus
link the HOMFLY polynomial has the form $(m>0)$:
$$
X^{(n,m)}={\lambda^{{1\over 2}(n-1)(m-1)}\over (t-1)^{s-1}} P^{(n,m)},
\eqn\guess
$$
where $s$ is the greatest common divisor of $n$ and $m$, \ie, the number of
components of the link, and $P^{(n,m)}$ is a true polynomial (no negative
powers of $t$ and $\lambda$ appear in $P^{(n,m)}$) in $t$ and $\lambda$. The
exponent of $\lambda$ in \guess\ has a special significance for torus knots. It
is called the {\it Gordian number} and according to Milnor's conjecture it
corresponds to the unknotting number of the knot, \ie, the minimun number of
uncrossings that have to performed on a knot to make it the unknot. Notice that
for torus knots $n$ and $m$ are coprime and then the Gordian number is always
an integer. If the conjecture
\guess\ is correct, it seems that the Gordian number might play also an
important role for torus links.
One would like to prove \guess\ from \cocotier. We leave this
problem for future work. For the situation in which $n$ and $m$ are coprime,
the existence of the decomposition \guess\ was proved in [\poli].
In case that the decomposition
shown in \guess\ holds for torus links  one would like to know if there are
similar decompositions as the one in \guess\ for a general link or, if not, for
which particular classes of links it might be valid.

The results presented in this work show how powerful Chern-Simons theory 
can be in
computing link invariants once the Hilbert space on the corresponding Riemann
surface is known. Unfortunately, it is not possible at the moment to go beyond
$g=1$ though recent work  \REF\gawd{K. Gawedzki, ``$SU(2)$ WZW at higher genera
from gauge field functional integral", hep-th/9312051} [\gawd] seems to provide
a promising  framework in this respect. Further work in this direction is
needed.

 Another important fact is that the structure 
of the computations carried out in sect.
3 has strong similarities with the one performed in [\rosso] using quantum
groups. This observation should be pushed further since it might reveal a more
explicit connection between quantum groups and Chern-Simons gauge theory
\REF\moore{G. Moore and N. Reshetikhin\journal\np&B328(89)557} [\moore].

In the study carried out in sect. 5 on the limit $N\rightarrow 0$ to obtain the
Alexander  polynomial it was observed that such a polynomial is obtained when
considering the normalized invariant. The natural invariant (no normalized by
the unknot) generated by Chern-Simons gauge theory vanishes as one would
naively expect. It would be very interesting to understand at the action level
the reason behind this behavior. In particular, it would be very interesting if
following this correspondence one could find new relations between Chern-Simons
gauge theory and other models leading to the Alexander polynomials as the ones
in  [\roz] and \REF\kauf{L.H. Kauffman and H. Saleur\journal\cmp&141(91)293}
[\kauf].

\vskip1cm

\ack
We would like to thank A. V. Ramallo for very helpful
dicussions. 
This work was supported in part by DGICYT under grant PB90-0772 and
by CICYT under grant AEN94-0928.

\endpage
\Appendix{A}

\section{Group-theoretical Conventions.}

In this subsection of the  Appendix we will summarize our conventions
for the group $SU(N)$.
The $N^2 -1$ generators of  $SU(N)$ are chosen
antihermitian and in the fundamental representation are normalized 
as follows: 
$$
\tr(T^i T^j)=-\delta^{ij}.
\eqn\ai
$$
The rank of $SU(N)$ is $r=N-1$. There are $r$ fundamental roots which 
will be denoted by
$\alpha_i$, $i=1,...,r$. They are chosen to have length $\sqrt{2}$,
${\alpha_i}^2=2$, and the following Cartan matrix, 
$g_{ij}=\alpha_i\cdot\alpha_j$,
$$
g_{ij}=\left(\matrix{2&-1&0&0&\cdot&\cdot&\cdot&\cdot&0\cr
                -1&2&-1&0&\cdot&\cdot&\cdot&\cdot&0\cr
                0&-1&2&-1&\cdot&\cdot&\cdot&\cdot&0\cr
                \cdot&\cdot&\cdot&\cdot&\cdot&\cdot&\cdot&\cdot&\cdot\cr
                \cdot&\cdot&\cdot&\cdot&\cdot&\cdot&\cdot&\cdot&\cdot\cr
                0&0&0&0&\cdot&\cdot&-1&2&-1\cr
                0&0&0&0&\cdot&\cdot&0&-1&2\cr}\right)
\eqn\aii
$$
We will denote the root lattice by $\rl$.
 This $r$-dimensional space is generated by the fundamental roots $\alpha_i$,
which can be taken as a basis, the root basis. Any vector $x$  in this basis
has components ${\hat x}_i$ given by: $$ x=\sum_{i=1}^{r} {\hat x}_i\alpha_i.
\eqn\aiip
$$
Among all the roots in $\rl$ there is a subset which plays an important role in
the calculation performed in the paper. These are the positive roots. 
For $SU(N)$ they take the form,
$$
\alpha_i + \alpha_{i+1} +\cdots +\alpha_j,\,\,\,\,\,\  1 \le
i\le j  \le r.
\eqn\posroots
$$

The fundamental weights $\lambda_i,$ $i=1,...,r,$ 
are defined by
$$
2\,{\alpha_i\cdot\lambda_j
\over{\alpha_i\cdot\alpha_i}}=\delta_{ij},
\eqn\aiv
$$
which in the case of $SU(N)$, since the squared length of the fundamental roots
is 2, becomes
$$
\alpha_i\cdot\lambda_j=\delta_{ij},
\eqn\av
$$
\ie, the fundamental weights are the duals to the fundamental
roots. The fundamental weights generate an $r$-dimensional
lattice over ${\bf Z}$  called the weight lattice which will be denoted by
$\wl$. The lattices $\rl$ and $\wl$ are dual to each other and
$\rl\subset\wl$. 
The $r$-dimensional basis spanned by the fundamental weights is
called the Dynkin basis. Any vector $x$ has in this 
basis components $x_i$ given by:
$$
x=\sum_{i=1}^{r}  x_i\lambda_i.
\eqn\avi
$$
The matrix $(g^{-1})_{ij}=\lambda_i\cdot\lambda_j$ is the inverse of the
Cartan matrix, $(g^{-1})_{ij}\cdot
g_{jk}=\delta_{ik}$, and turns out to be
$$
(g^{-1})_{ij}={\hbox{\rm Inf}}\{i,j\}- {ij\over N}.
\eqn\avii 
$$
Among the weights in $\wl$ there is one which plays an important role in
Chern-Simons theory because it can be regarded as the vacuum. This weight
is denoted by $\rho$ and all its components are one:
$$
\rho = \sum_{i=1}^r \lambda_i.
\eqn\elrho
$$

The irreducible representations of $SU(N)$ are characterized by 
highest weights  $\Lambda$. Highest weights can
be written uniquely as a linear combination of fundamental weights
with non-negative integer coefficients $h_i$,
$$
\Lambda = \sum_{i=1}^r  h_i\lambda_i.
\eqn\aix
$$
The set of weights of an irreducible representation of highest weight $\Lambda$
will be denoted as $M_\Lambda$. To build this set one may use   the following
rule:
 if a weight
$\mu\in M_\Lambda$  has the $k^{th}$ Dynkin component
greater than zero (\ie, ${\mu}_k>0$), then the vectors obtained
by subtracting  $t\alpha_k$ ($t=1,...,{\mu}_k$) from $\mu$ 
are also elements of $M_\Lambda$. One can start applying this rule to $\Lambda$
and then to the successive weights to build the different elements of
$M_\Lambda$. The multiplicities of each weight can be obtained using the
Freudenthal formula \REF\book{N. Jacobson, {\it Lie Algebras,}
Wiley-Interscience, New York, 1962; J. E. Humphreys, {\it Introduction to Lie
Algebras and Representation Theory}, Springer, New York, 1972.} [\book].

The Weyl group is generated by $r$ reflections $\sigma_i$, $i=1,...,r$, on
weight space
$$
x\in \wl, \,\,\,\,\,\, \sigma_i(x) = x - \alpha_i(\alpha_i\cdot x).
\eqn\apnu
$$
It divides the weight lattice $\wl$ into $N!$ factorial
domains. The fundamental domain or Weyl chamber is chosen to be the one
containing all the weights $x\in\wl$ such that
$$
\alpha_i\cdot x=x_i\geq 0.
\eqn\axvi
$$

The Weyl  character for an irreducible representation of highest weight
$\Lambda$ is defined as
$$
\ch_\Lambda(a)=\sum_{\mu\in M_\Lambda} \ex^{a\cdot\mu},
\eqn\apuno
$$
where $a=a_i\lambda^i$. The Weyl character satisfies the equation
[\book],
$$
\ch_\Lambda(a)={\sum_{w\in W}\epsilon(w) \ex^{w(\Lambda+\rho)\cdot a}
\over \sum_{w\in W}\epsilon(w) \ex^{w(\rho)\cdot a} },
\eqn\apdos
$$
known as the Weyl character formula. When $a=-\rho$,
we have an expression for the character \REF\kac{V. G.
Kac, {\it Infinite-dimensional Lie algebras}, Birkh\"auser, Boston,
1983.} 
[\kac] which is particularly useful:
$$
\sum_{\mu \in M_\Lambda} \ex^{-\mu \cdot \rho} =
\prod_{\alpha>0}{\ex^{{1\over 2}\alpha\cdot(\rho + 
\Lambda)}-\ex^{-{1\over 2}\alpha \cdot(\rho+\Lambda)}\over \ex^{{1\over 2}
\alpha\cdot\rho}-\ex^{-{1\over 2}\alpha\cdot\rho}}.
\eqn\lechuguino
$$
where the product is performed over the set of positive roots \posroots.

An important set of weights used in this work is the one made by
Weyl-antisymmetric combinations of weights in $\wl / l\rl$ where $l=k+N$ and
$k$ is an arbitrary non-negative integer. This set of weights builds the
fundamental chamber ${\cal F}_l$ and is made up with weights of the form
$x=x_i\lambda_i$ where,
$$
x_i>0, \,\,\,\,\,\,\,\,\,\,\,\,\,\,\,\,
\sum_{i=1}^r x_i < l.
\eqn\funch
$$
The set ${\cal F}_l$ has a total of $(l-1)!/(l-N)!(N-1)!$ elements.

\section {Fundamental representation of $SU(N)$}
In this subsection we present results concerning the fundamental 
representation of
$SU(N)$. The 
fundamental representation of $SU(N)$ is associated to the highest weight
$\Lambda = \lambda_1 = (1,0,\cdots,0)$, and the corresponding weight space is:
$$ M_{\lambda_1}=\{ \mu_i: 1\le i \le N \}, $$
where:
$$
\eqalign{
\mu_1=&\lambda_1,\cr
\mu_2=& \lambda_1-\alpha_1 = (-1,1,0,\cdots,0),\cr
\mu_3=& \mu_2-\alpha_2 = (0,-1,1,0,\cdots,0),\cr
&\,\,\,\,\,\,\,\,\,\,\,\,\,\,\,\,\,\ \vdots \cr
      \mu_N =&\mu_{N-1}-\alpha_{N-1} = (0,\cdots,0,-1).\cr}
\eqn\leila
$$
  We can write these weights as follows:
$$ 
\mu_i = \sum_{j=1}^{N-1} \mu^{j}_i \lambda_j,
\eqn\mortadelo
$$
where 
$$
\mu^{j}_i =\cases {1, & if $i=j$,\cr
                   -1, & if $i-1=j$,\cr
                    0, & otherwise.\cr}
\eqn\carpanta
$$   
 We also need the scalar products $\rho\cdot\mu_i$,
$\mu_i\cdot\mu_j$. Using the form \mortadelo\ and \avii, one easily finds:
$$
\mu_i\cdot\mu_j=\cases { 1-{1 \over N}, & if $i=j$ \cr
                      -{1 \over N}, & if  $i \not= j$ \cr}
\eqn\itamara
$$
In a similar fashion one obtains:
$$
\rho\cdot\mu_i ={1 \over 2}(N-2i+1),\,\,\,\,\ 1\le i \le N.
\eqn\peixoto
$$
The action of the Weyl reflections on the fundamental weights 
$\lambda_i$ follows from \apnu:
$$
 \eqalign{
\sigma_i(\lambda_j)=&0, \,\,\,\,\ i \not = j,\cr
\sigma_i(\lambda_i)=&\lambda_i-\alpha_i=\lambda_{i-1}+\lambda_{i+1}-\lambda_i,
\,\,\,\ 1<i<N, \cr
 \sigma_1(\lambda_1)=&\lambda_2-\lambda_1,\cr
\sigma_{N-1}(\lambda_{N-1})=&\lambda_{N-2}-\lambda_{N-1}.\cr}
\eqn\tatiana
$$

\endpage
\Appendix{B}

In this Appendix we will prove the relation \pesa\ used in sect. 5. To prove it
we need to be more specific about the way we have been labeling ordered
partitions so far. The ordered partitions $(k_\lambda)$ of $s$ will be labeled
by $N_s$ vectors
${\bf P}^s_j$, $j=1,\cdots,N_s$. Each vector ${\bf P}^s_j$ has $r_j^s$
components $({\bf P}^s_j)_\lambda$, $\lambda=1,\cdots,r_j^s$. First, we will
derive a formula for the number of ordered partitions of $s$.
We begin by pointing out that the ordered partitions of $s$ can be listed in
the following way:
$$
\eqalign{& (s), \cr
&(s-1,{\bf P}^1_j), \cr
&(s-2,{\bf P}^2_j), \cr
&\,\,\,\,\,\,\,\,\,\,\,\,\,\vdots \cr
&(s-i,{\bf P}^i_j), \cr
&\,\,\,\,\,\,\,\,\,\,\,\,\,\vdots \cr
&(1,{\bf P}^{s-1}_j), \cr}
\qquad\qquad
\eqalign{\cr
j=1,\cdots,N_1, \cr
j=1,\cdots,N_2, \cr
&\hbox{\hskip-1.5cm}\vdots \cr
j=1,\cdots,N_i, \cr
&\hbox{\hskip-1.5cm}\vdots \cr
j=1,\cdots,N_{s-1}. \cr}
\eqn\yale
$$
According to this, the number of partitions of $s$, $N_s$, satisfies the
following relation:
$$
N_s= 1 + \sum_{i=1}^{s-1} N^i,
\eqn\leya
$$
and $N_1=1$. Let us prove  inductively that $N_s= 2^{s-1}$. Let us assume that
for $i<s$, $N_i=2^{i-1}$. Making use of \leya\ to compute $N_s$, one finds,
$$
N_s= 1 + \sum_{i=1}^{s-1} 2^{i-1} = 2^{s-1}.
\eqn\eneese
$$

To prove \pesa\ let us begin defining the following function:
$$
F(s)=\cases{ \sum_{j=1}^{N_s} {s!\over \prod_{\lambda=1}^{r_j^s}
({\bf P}_j^s)_\lambda ! } (-1)^{r_j^s}, &if $s\ge 1$,\cr
-1,& if $s = 1$.\cr}
\eqn\laefe
$$
Organizing the ordered partitions of $s$ as in \yale, this function satisfies
the following property:
$$
\eqalign{
F(s)=& -1-\sum_{i=1}^{s-1}\sum_{j=1}^{N_i}
{s!\over (s-i)! \prod_{\lambda=1}^{r_j^i} ({\bf P}_j^i)_\lambda !}
(-1)^{r_j^i}\cr
=& -1-\sum_{i=1}^{s-1}{s!\over (s-i)! i!} F(i).\cr}
\eqn\vieira
$$
We will prove now by induction that,
 $$
F(s)=(-1)^s.
\eqn\vieiro
$$
 Let us assume that  $F(i)=(-1)^i$ for 
$i<s$. From \vieira\ one finds,
$$
F(s)=-1-\sum_{i=1}^{s-1} {s!\over (s-i)! i!}(-1)^i =
(-1)^s + (1-1)^s = (-1)^s.
\eqn\concha
$$

Finally, we will consider the sum:
$$
f_s(x)=\sum_{j=1}^{N_s} {s!\over \prod_{\lambda=1}^{r_j^s}
({\bf P}_j^s)_\lambda ! } (-1)^{r_j^s+1}(x^{({\bf P}_j^s)_1}-1),
\eqn\conce
$$
which is a quantity of the type entering \pesa.
Making the same rearrangement of the partitions of $s$ as the one done in
\yale, and using \vieiro:
$$
\eqalign{
f_s(x) = & x^s-1+\sum_{i=1}^{s-1}\sum_{j=1}^{N_i}
{s!\over (s-i)! \prod_{\lambda=1}^{r_j^i} ({\bf P}_j^i)_\lambda ! }
(-1)^{r_j^i}(x^{s-i}-1) \cr
= & x^s-1+\sum_{i=1}^{s-1} {s!\over (s-i)! i!} F(i) (x^{s-i}-1)\cr
= & \sum_{i=0}^{s} {s!\over (s-i)! i!} (-1)^i (x^{s-i}-1) = (x-1)^s, \cr}
\eqn\laecu
$$
which is the relation used in \pesa.

\endpage

\Appendix{C}

In this appendix we present a computer routine written in
{\sl Mathematica}$^{\hbox{\sevenrm TM}}$  which implements
the formula for the HOMFLY polynomial for an arbitrary torus
link $(n,m)$ obtained in theorem 4.2. The input data for this routine 
are two integers {\tt n} and {\tt m} with ${\tt n}>0$.
The variable $\lambda$ of the link invariant has been denoted by
{\tt lb} in the computer routine.
 The routine is divided in 
blocks. After the initialization, the first block generates the 
ordered partitions of {\tt s}, being {\tt s} the greatest common divisor of
{\tt n} and {\tt m}. For each ordered partition, the arrangements of indices
which are selected in theorem 4.1 are built in the next block. The quantity
{\tt dim2} constitutes  an upper bound for the number of arrangements selected
or, equivalently, the number of weights which contribute, which are generated
in the next block making use again of the result stated in theorem 4.1.
Finally, in the last block the formula presented in theorem 4.2 is
computed and the result is printed out. The routine prints before a list
of all the weights in ${\cal F}_l$ which contribute for each ordered
partition as well as the corresponding sign due to the Weyl transformation
carried out. If this intermediate listing is not desired it can be omited
commenting out the last two lines of the block previous to the last one.
We end this appendix with a printout of the computer routine.

\vskip0.7cm

\epsfbox{part1.eps} 
\endpage
\epsfbox{part2.eps}
\endpage
\epsfbox{part3.eps} 

\endpage

\Appendix{D}

In this Appendix we present a list of the HOMFLY polynomial for
some selected torus links. All of them have been computed using the
implementation of  formula (4.9) in the computer routine presented in
the previous Appendix. 

$$
\eqalign{
X^{(2,3)}=&\lambda (1+t^2-\lambda t^2) \cr
X^{(2,2)}=&\lambda^{1\over 2}\,(t-1)^{-1}
        (-1+t-t^2 + \lambda t^2) \cr
X^{(2,4)}=&{{{{\lambda}}^{{3\over 2}}}\,(t-1)^{-1}
     \left( -1 + t - {t^2} + {\lambda}\,{t^2} + {t^3} - 
       {\lambda}\,{t^3} - {t^4} + {\lambda}\,{t^4} \right) }\cr
X^{(3,6)}=&{{{\lambda}}^5}\,{{{\left( t-1 \right) }^{-2}}}
       ( 1 - 2\,t + 2\,{t^2} - 
       {\lambda}\,{t^2} - {t^3} + {\lambda}\,{t^3} - 
       {\lambda}\,{t^5} + {{{\lambda}}^2}\,{t^5} + {t^6} + 
       {\lambda}\,{t^6} \cr &
      - 2\,{{{\lambda}}^2}\,{t^6} - 
       2\,{\lambda}\,{t^7} + 2\,{{{\lambda}}^2}\,{t^7} + 
       {\lambda}\,{t^8} - {{{\lambda}}^2}\,{t^8} - {t^9} - 
       {\lambda}\,{t^9} + 2\,{{{\lambda}}^2}\,{t^9} + 
       2\,{t^{10}} \cr &
     - 2\,{{{\lambda}}^2}\,{t^{10}} - 
       2\,{t^{11}} + {\lambda}\,{t^{11}} + 
       {{{\lambda}}^2}\,{t^{11}} + {t^{12}} - 
       {\lambda}\,{t^{12}} ) \cr
X^{(4,6)}=&{{{\lambda}}^{{{15}\over 2}}}\, (t-1)^{-1}
       ( -1 + t - {t^2} + {\lambda}\,{t^2} - {t^4} + 
       {\lambda}\,{t^4} + {t^5} - {{{\lambda}}^2}\,{t^5} - 
       2\,{t^6} + 2\,{\lambda}\,{t^6} \cr &
     + 2\,{t^7}-{\lambda}\,{t^7} - {{{\lambda}}^2}\,{t^7} - 2\,{t^8} + 
       2\,{\lambda}\,{t^8} + 2\,{t^9} - 2\,{\lambda}\,{t^9} - 
       {{{\lambda}}^2}\,{t^9} + {{{\lambda}}^3}\,{t^9} - 
       2\,{t^{10}} \cr &
     + 2\,{\lambda}\,{t^{10}} + 
       {{{\lambda}}^2}\,{t^{10}} - {{{\lambda}}^3}\,{t^{10}} + 
       {t^{11}} - {\lambda}\,{t^{11}} - 
       {{{\lambda}}^2}\,{t^{11}} + {{{\lambda}}^3}\,{t^{11}} - 
       {t^{12}} + 2\,{\lambda}\,{t^{12}} \cr &
     - {{{\lambda}}^3}\,{t^{12}} - {{{\lambda}}^2}\,{t^{13}} + 
       {{{\lambda}}^3}\,{t^{13}} - {t^{14}} + 
       {\lambda}\,{t^{14}} + {t^{15}} - 
       {{{\lambda}}^2}\,{t^{15}} - {t^{16}} + 
       {\lambda}\,{t^{16}} ) \cr
X^{(4,8)}=&{{\lambda}}^{21\over 2}\,(t-1)^{-3}
      ( -1 + 3\,t - 4\,{t^2} + {\lambda}\,{t^2} + 
       3\,{t^3} - 2\,{\lambda}\,{t^3} - 2\,{t^4} + 
       2\,{\lambda}\,{t^4} + 3\,{t^5} \cr & 
    -  2\,{\lambda}\,{t^5} 
       -{{{\lambda}}^2}\,{t^5} - 5\,{t^6} + 
       3\,{\lambda}\,{t^6} + 2\,{{{\lambda}}^2}\,{t^6} + 
       6\,{t^7} - 4\,{\lambda}\,{t^7} - 
       2\,{{{\lambda}}^2}\,{t^7} - 6\,{t^8} \cr &
     + 4\,{\lambda}\,{t^8} + 2\,{{{\lambda}}^2}\,{t^8} + 
       3\,{t^9} - {\lambda}\,{t^9} - 
       3\,{{{\lambda}}^2}\,{t^9} + {{{\lambda}}^3}\,{t^9} - 
       {\lambda}\,{t^{10}} + 4\,{{{\lambda}}^2}\,{t^{10}} \cr &
     - 3\,{{{\lambda}}^3}\,{t^{10}} + 3\,{\lambda}\,{t^{11}} - 
       7\,{{{\lambda}}^2}\,{t^{11}} + 
       4\,{{{\lambda}}^3}\,{t^{11}} - {t^{12}} - 
       3\,{\lambda}\,{t^{12}} + 7\,{{{\lambda}}^2}\,{t^{12}} \cr &
    -  3\,{{{\lambda}}^3}\,{t^{12}} + 3\,{\lambda}\,{t^{13}} - 
       5\,{{{\lambda}}^2}\,{t^{13}} + 
       2\,{{{\lambda}}^3}\,{t^{13}} - 3\,{\lambda}\,{t^{14}} + 
       3\,{{{\lambda}}^2}\,{t^{14}} + 3\,{t^{15}} \cr &
    +  3\,{\lambda}\,{t^{15}} - 5\,{{{\lambda}}^2}\,{t^{15}} - 
       {{{\lambda}}^3}\,{t^{15}} - 6\,{t^{16}} - 
       {\lambda}\,{t^{16}} + 7\,{{{\lambda}}^2}\,{t^{16}} + 
       6\,{t^{17}} - {\lambda}\,{t^{17}} \cr &
    -  7\,{{{\lambda}}^2}\,{t^{17}} + 
       2\,{{{\lambda}}^3}\,{t^{17}} - 5\,{t^{18}} +
       4\,{\lambda}\,{t^{18}} + 4\,{{{\lambda}}^2}\,{t^{18}} - 
       3\,{{{\lambda}}^3}\,{t^{18}} + 3\,{t^{19}} - 
       4\,{\lambda}\,{t^{19}} \cr &
     - 3\,{{{\lambda}}^2}\,{t^{19}} + 
       4\,{{{\lambda}}^3}\,{t^{19}} - 2\,{t^{20}} + 
       3\,{\lambda}\,{t^{20}} + 2\,{{{\lambda}}^2}\,{t^{20}} - 
       3\,{{{\lambda}}^3}\,{t^{20}} + 3\,{t^{21}} - 
       2\,{\lambda}\,{t^{21}} \cr &
     - 2\,{{{\lambda}}^2}\,{t^{21}} + 
       {{{\lambda}}^3}\,{t^{21}} - 4\,{t^{22}} + 
       2\,{\lambda}\,{t^{22}} + 2\,{{{\lambda}}^2}\,{t^{22}} + 
       3\,{t^{23}} - 2\,{\lambda}\,{t^{23}} - 
       {{{\lambda}}^2}\,{t^{23}} \cr &
     - {t^{24}} + 
       {\lambda}\,{t^{24}} ) \cr }
$$

$$
\eqalign{
X^{(6,9)}=&{{{{\lambda}}^{20}}\,{{{\left( t-1 \right) }^{-2}}}}
       ( 1 - 2\,t + 2\,{t^2} - 
       {\lambda}\,{t^2} - {t^3} + {\lambda}\,{t^3} + {t^4} - 
       {\lambda}\,{t^4} - {t^5} + {{{\lambda}}^2}\,{t^5} \cr & 
      +2\,{t^6} - {\lambda}\,{t^6} - {{{\lambda}}^2}\,{t^6} - 
       3\,{t^7} + {\lambda}\,{t^7} + 
       2\,{{{\lambda}}^2}\,{t^7} + 4\,{t^8} - 
       3\,{\lambda}\,{t^8} - {{{\lambda}}^2}\,{t^8} - 
       3\,{t^9} \cr &
     + 2\,{\lambda}\,{t^9} + 
       2\,{{{\lambda}}^2}\,{t^9} - {{{\lambda}}^3}\,{t^9} + 
       2\,{t^{10}} - 2\,{\lambda}\,{t^{10}} - 
       {{{\lambda}}^2}\,{t^{10}} + {{{\lambda}}^3}\,{t^{10}} - 
       2\,{t^{11}} \cr &
     + 4\,{{{\lambda}}^2}\,{t^{11}} - 
       2\,{{{\lambda}}^3}\,{t^{11}} + 4\,{t^{12}} - 
       {\lambda}\,{t^{12}} - 4\,{{{\lambda}}^2}\,{t^{12}} + 
       {{{\lambda}}^3}\,{t^{12}} - 6\,{t^{13}} + 2\,{\lambda}\,{t^{13}} \cr & 
     + 6\,{{{\lambda}}^2}\,{t^{13}} - 
       2\,{{{\lambda}}^3}\,{t^{13}} + 6\,{t^{14}} - 
       4\,{\lambda}\,{t^{14}} - 4\,{{{\lambda}}^2}\,{t^{14}} + 
       {{{\lambda}}^3}\,{t^{14}} + {{{\lambda}}^4}\,{t^{14}} - 
       4\,{t^{15}} \cr & 
     + 4\,{\lambda}\,{t^{15}} + 
       4\,{{{\lambda}}^2}\,{t^{15}} - 
       3\,{{{\lambda}}^3}\,{t^{15}} - 
       {{{\lambda}}^4}\,{t^{15}} + 2\,{t^{16}} - 
       2\,{\lambda}\,{t^{16}} - 4\,{{{\lambda}}^2}\,{t^{16}} + 
       3\,{{{\lambda}}^3}\,{t^{16}}  \cr &  
     + {{{\lambda}}^4}\,{t^{16}} - 2\,{\lambda}\,{t^{17}} + 
       6\,{{{\lambda}}^2}\,{t^{17}} - 
       4\,{{{\lambda}}^3}\,{t^{17}} + 2\,{t^{18}} - 
       6\,{{{\lambda}}^2}\,{t^{18}} + 
       3\,{{{\lambda}}^3}\,{t^{18}} + 
       {{{\lambda}}^4}\,{t^{18}} \cr & 
     - 2\,{t^{19}} - 
       2\,{\lambda}\,{t^{19}} + 8\,{{{\lambda}}^2}\,{t^{19}} - 
       2\,{{{\lambda}}^3}\,{t^{19}} - 
       2\,{{{\lambda}}^4}\,{t^{19}} + 2\,{t^{20}} - 
       2\,{\lambda}\,{t^{20}} - 2\,{{{\lambda}}^2}\,{t^{20}} \cr &  
     + 3\,{{{\lambda}}^4}\,{t^{20}} - 
       {{{\lambda}}^5}\,{t^{20}} - 2\,{\lambda}\,{t^{21}} + 
       6\,{{{\lambda}}^2}\,{t^{21}} - 
       3\,{{{\lambda}}^3}\,{t^{21}} - 
       3\,{{{\lambda}}^4}\,{t^{21}} + 
       2\,{{{\lambda}}^5}\,{t^{21}} \cr & 
     + 2\,{t^{22}} - 
       2\,{\lambda}\,{t^{22}} + 2\,{{{\lambda}}^4}\,{t^{22}} - 
       2\,{{{\lambda}}^5}\,{t^{22}} - 2\,{t^{23}} - 
       2\,{\lambda}\,{t^{23}} + 6\,{{{\lambda}}^2}\,{t^{23}} - 
       4\,{{{\lambda}}^3}\,{t^{23}} \cr &  
     + {{{\lambda}}^4}\,{t^{23}} + {{{\lambda}}^5}\,{t^{23}} + 
       2\,{t^{24}} - 2\,{\lambda}\,{t^{24}} - 
       {{{\lambda}}^3}\,{t^{24}} + {{{\lambda}}^4}\,{t^{24}} - 
       2\,{\lambda}\,{t^{25}} + 6\,{{{\lambda}}^2}\,{t^{25}} \cr &
     - 4\,{{{\lambda}}^3}\,{t^{25}} + 2\,{t^{26}} - 
       2\,{{{\lambda}}^2}\,{t^{26}} + 
       {{{\lambda}}^4}\,{t^{26}} - {{{\lambda}}^5}\,{t^{26}} - 
       4\,{t^{27}} - 2\,{\lambda}\,{t^{27}} + 
       8\,{{{\lambda}}^2}\,{t^{27}} \cr & 
     - 3\,{{{\lambda}}^3}\,{t^{27}} + 
       {{{\lambda}}^4}\,{t^{27}} + 6\,{t^{28}} - 
       2\,{\lambda}\,{t^{28}} - 6\,{{{\lambda}}^2}\,{t^{28}} + 
       2\,{{{\lambda}}^4}\,{t^{28}} - 6\,{t^{29}} + 
       4\,{\lambda}\,{t^{29}} \cr & 
     + 6\,{{{\lambda}}^2}\,{t^{29}} - 
       2\,{{{\lambda}}^3}\,{t^{29}} - 
       3\,{{{\lambda}}^4}\,{t^{29}} + 
       {{{\lambda}}^5}\,{t^{29}} + 4\,{t^{30}} - 
       4\,{\lambda}\,{t^{30}} - 4\,{{{\lambda}}^2}\,{t^{30}} \cr &  
     + 3\,{{{\lambda}}^3}\,{t^{30}} + 
       3\,{{{\lambda}}^4}\,{t^{30}} - 
       2\,{{{\lambda}}^5}\,{t^{30}} - 2\,{t^{31}} + 
       2\,{\lambda}\,{t^{31}} + 4\,{{{\lambda}}^2}\,{t^{31}} - 
       4\,{{{\lambda}}^3}\,{t^{31}} \cr &  
     - 2\,{{{\lambda}}^4}\,{t^{31}} + 
       2\,{{{\lambda}}^5}\,{t^{31}} + 2\,{t^{32}} - 
       {\lambda}\,{t^{32}} - 4\,{{{\lambda}}^2}\,{t^{32}} + 
       3\,{{{\lambda}}^3}\,{t^{32}} + 
       {{{\lambda}}^4}\,{t^{32}} - {{{\lambda}}^5}\,{t^{32}} \cr &  
     - 3\,{t^{33}} + 6\,{{{\lambda}}^2}\,{t^{33}} - 
       3\,{{{\lambda}}^3}\,{t^{33}} + 4\,{t^{34}} - 
       2\,{\lambda}\,{t^{34}} - 4\,{{{\lambda}}^2}\,{t^{34}} + 
       {{{\lambda}}^3}\,{t^{34}} + {{{\lambda}}^4}\,{t^{34}} \cr &
     - 3\,{t^{35}} + 2\,{\lambda}\,{t^{35}} + 
       4\,{{{\lambda}}^2}\,{t^{35}} - 
       2\,{{{\lambda}}^3}\,{t^{35}} - 
       {{{\lambda}}^4}\,{t^{35}} + 2\,{t^{36}} - 
       3\,{\lambda}\,{t^{36}} - {{{\lambda}}^2}\,{t^{36}} \cr &  
     + {{{\lambda}}^3}\,{t^{36}} + {{{\lambda}}^4}\,{t^{36}} - 
       {t^{37}} + {\lambda}\,{t^{37}} + 
       2\,{{{\lambda}}^2}\,{t^{37}} - 
       2\,{{{\lambda}}^3}\,{t^{37}} + {t^{38}} - 
       {\lambda}\,{t^{38}} - {{{\lambda}}^2}\,{t^{38}} \cr & 
     + {{{\lambda}}^3}\,{t^{38}} - {t^{39}} + 
       2\,{{{\lambda}}^2}\,{t^{39}} - 
       {{{\lambda}}^3}\,{t^{39}} + 2\,{t^{40}} - 
       {\lambda}\,{t^{40}} - {{{\lambda}}^2}\,{t^{40}} - 
       2\,{t^{41}} + {\lambda}\,{t^{41}} \cr &  
     + {{{\lambda}}^2}\,{t^{41}} + {t^{42}} - 
       {\lambda}\,{t^{42}}) 
\cr}
$$
\endpage
\refout
\end